\documentclass[11pt,letterpaper]{article}

\usepackage[letterpaper,margin=1in]{geometry}

\usepackage{amsmath,amssymb}
\usepackage{thmtools}
\usepackage{amsthm}
\usepackage{enumitem}
\usepackage{booktabs}

\usepackage[hidelinks,hypertexnames=false]{hyperref}
\usepackage[nameinlink,capitalize]{cleveref}

\usepackage{tikz}
\usetikzlibrary{positioning,fit,backgrounds}

\usepackage[mathlines]{lineno}

\newtheoremstyle{compact}
  {6pt}          
  {6pt}          
  {\itshape}
  {}
  {\bfseries}
  {.}
  {.5em}
  {}

\theoremstyle{compact}

\newtheorem{theorem}{Theorem}[section]
\newtheorem{lemma}[theorem]{Lemma}

\theoremstyle{definition}
\newtheorem{definition}[theorem]{Definition}

\crefname{appendix}{Appendix}{Appendices}

\AddToHook{cmd/appendix/before}{%
  \crefalias{section}{appendix}%
  \crefalias{subsection}{appendix}%
}

\title{Faster Exact Algorithms for 3- and 4-Coloring: Deterministic Improvements and Further Randomized Speedups}

\author{
  Ryosuke Yamano\thanks{
Department of Computer Science, Graduate School of Information Science and Technology, The University of Tokyo, Japan.
ryoyamano15@g.ecc.u-tokyo.ac.jp.
}
  \and
  Tetsuo Shibuya\thanks{
Division of Medical Data Informatics, Human Genome Center, Institute of Medical Science, The University of Tokyo, Japan.
tshibuya@hgc.jp.
Supported by KAKENHI Grant Number 23K28035.
}
}

\date{}

\newcommand{\ThreeDetBase}{1.3204}
\newcommand{\ThreeRandBase}{1.3159}
\newcommand{\FourDetBase}{1.6975}
\newcommand{\FourRandBase}{1.6933}

\begin{document}

\maketitle
\thispagestyle{empty}

{
\let\small\normalsize
\begin{abstract}
We study exact exponential-time algorithms for 3- and 4-coloring. For 3-coloring, the best previously published running-time bound is $O^*(1.3289^n)$, due to Eppstein (SODA 2001). We improve this bound to $O^*(\ThreeDetBase^n)$ deterministically and $O^*(\ThreeRandBase^n)$ using randomization. For 4-coloring, the best previously known running-time bound is $O^*(1.7159^n)$, due to Wu, Gu, Jiang, Shao, and Xu (ESA 2024). We improve this bound to $O^*(\FourDetBase^n)$ deterministically and $O^*(\FourRandBase^n)$ using randomization.

Our algorithms for both problems combine constraint satisfaction algorithms with dynamic programming over tree and path decompositions, exploiting their complementary strengths on different graph structures. For 6-coloring, the best published randomized running-time bound is exponentially smaller than the best published deterministic bound. For 3- and 4-coloring, however, no such improvement through randomization was previously known. To the best of our knowledge, our results provide the first randomized running-time bounds for these two problems that exponentially improve upon the best previously published deterministic bounds. This raises the question of whether randomization offers an inherent exponential advantage for 3- and 4-coloring.
\end{abstract}
}

\clearpage

%

\pagenumbering{gobble}
\tableofcontents
\clearpage

\pagenumbering{arabic}

\section{Introduction}

Given a graph $G = (V, E)$, a $k$-coloring is a function $f: V \rightarrow \{1, \ldots, k\}$ such that $f(u) \neq f(v)$ for every edge $uv \in E$.
The smallest $k$ for which $G$ admits a $k$-coloring is its \emph{chromatic number}.
Computing this number is the \emph{chromatic number problem}, whose decision version is one of Karp's 21 original NP-complete problems \cite{Karp1972}.
The best known polynomial-time approximation algorithm for the chromatic number achieves an approximation ratio of $O(n(\log\log n)^2/(\log n)^3)$ \cite{HALLDORSSON199319}, whereas approximating it within a factor of $n^{1-\varepsilon}$ is NP-hard for every constant $\varepsilon>0$ \cite{zuckerman2006linear}.

The $k$-coloring problem asks if a graph admits a $k$-coloring and is NP-hard for every fixed $k \geq 3$ \cite{lovasz1973coverings}.
Graph coloring has been widely studied through exact \cite{brown1972chromatic,sewell1996improved,FEDER2002192,bjorklund2008exact,BJORKLUND20111033,gaspers2023faster} and heuristic algorithms \cite{brelaz1979new,HERTZ20082551,malaguti2008metaheuristic}; see \cite{malaguti2010survey,MOSTAFAIE2020104850} for surveys.
We focus on exact algorithms.
On $n$-vertex graphs, $k$-coloring can be solved in $O^*(2^n)$ time\footnote{The $O^*(\cdot)$ notation suppresses polynomial factors.} \cite{bjorklund2006inclusion}.
For every fixed $k\ge7$, this is the best formally
published running-time bound
\footnote{
Preprints by Zamir and Pratt give randomized $O^*((2-\varepsilon_k)^n)$-time algorithms for every fixed $k$, with $\varepsilon_k>0$ \cite{zamir2026kcoloringfastercomputingchromatic,pratt2026breaking2nbarriergraph}.
Very recently, OpenAI and Anthropic released manuscripts claiming subexponential-time satisfiability algorithms \cite{OAI:The-Exponential-Time-Hypothesis-is-False-October-25-2026,Anthropic2026ETH}.
The claims by OpenAI and Anthropic remain unverified and are currently under examination, but, if correct, would imply subexponential-time algorithms for $k$-coloring for every fixed $k$.
}.
Interestingly, for $k \leq 6$, algorithms exponentially faster than $2^n$ are known.

For $k \in \{5,6\}$, Zamir broke the $2^n$ barrier by giving $O^*((2-\varepsilon)^n)$-time algorithms for some constant $\varepsilon>0$, deterministic for $k=5$ and randomized for $k=6$ \cite{zamir:LIPIcs.ICALP.2021.113}.
For $k \in \{3,4\}$, successive improvements have yielded faster algorithms.

For 3-coloring, the running-time bound improved from $O^*(1.4423^n)$ \cite{LAWLER197666} to $O^*(1.415^n)$ \cite{schiermeyer1993deciding}, $O^*(1.3446^n)$ \cite{beigel19953}, and $O^*(1.3289^n)$ \cite{eppstein2001improved}.
The last bound, obtained in 2001, remains the most recent formally published improvement
\footnote{
In a preprint, Meijer reports an
$O^*(1.3217^n)$-time algorithm for 3-coloring
\cite{meijer20233coloringtimeo13217n}.
However, we were unable to reproduce this bound from
the linear program presented in the preprint;
see \cref{app:meijer-lp} for details.
}.
For 4-coloring, the running-time bound improved from $O^*(1.8072^n)$ \cite{eppstein2001improved} to $O^*(1.7504^n)$ \cite{BYSKOV2004547}, $O^*(1.7272^n)$ \cite{fomin2007improved}, and $O^*(1.7159^n)$ \cite{wu_et_al:LIPIcs.ESA.2024.103}.
All of these algorithms for 3- and 4-coloring are deterministic.

We improve the bound for 3-coloring from $O^*(1.3289^n)$ \cite{eppstein2001improved} to $O^*(\ThreeDetBase^n)$ deterministically (\cref{thm:ThreeDet}) and $O^*(\ThreeRandBase^n)$ using randomization (\cref{thm:ThreeRand}).
For 4-coloring, we improve the bound from $O^*(1.7159^n)$ \cite{wu_et_al:LIPIcs.ESA.2024.103} to $O^*(\FourDetBase^n)$ deterministically (\cref{thm:FourDet}) and $O^*(\FourRandBase^n)$ using randomization (\cref{thm:FourRand}).
Our deterministic algorithms yield exponential improvements over previous bounds, and randomization yields further exponential improvements.

\begin{restatable}{theorem}{ThmThreeDet}
\label{thm:ThreeDet}
There is a deterministic algorithm for 3-Coloring on arbitrary
$n$-vertex graphs running in $O^*(\ThreeDetBase^n)$ time.
\end{restatable}

\begin{restatable}{theorem}{ThmThreeRand}
\label{thm:ThreeRand}
There is a randomized algorithm for 3-Coloring on arbitrary
$n$-vertex graphs running in $O^*(\ThreeRandBase^n)$ time.
\end{restatable}

\begin{restatable}{theorem}{ThmFourDet} \label{thm:FourDet}
There is a deterministic algorithm for 4-Coloring on arbitrary
$n$-vertex graphs running in $O^*(\FourDetBase^n)$ time.
\end{restatable}

\begin{restatable}{theorem}{ThmFourRand} \label{thm:FourRand}
There is a randomized algorithm for 4-Coloring on arbitrary $n$-vertex graphs running in $O^*(\FourRandBase^n)$ time.
\end{restatable}

Beyond improving running-time bounds, our results highlight the role of randomization in exact exponential-time graph coloring.
Whether randomization provides an inherent computational advantage is a fundamental question in complexity theory \cite{NISAN1994149,impagliazzo1997p,kabanets2003derandomizing,Vadhan2012}.
For exact exponential-time algorithms, this question takes a quantitative form: can randomization achieve an exponential base unattainable by deterministic algorithms?
For 6-coloring, the best known randomized running-time bound is exponentially smaller than the best known deterministic bound.
For 3- and 4-coloring, by contrast, the previous best bounds were achieved deterministically, and no improvement over them through randomization was known.
To the best of our knowledge, our results provide the first such improvements for these two problems.

Our randomized bounds are exponentially smaller even than the deterministic bounds established in this paper.
This raises a natural question: can the randomized bounds in \cref{thm:ThreeRand,thm:FourRand} also be achieved deterministically?
More broadly, does randomization offer an inherent exponential advantage for either problem?
Our results establish a gap between known upper bounds, leaving open whether it reflects a fundamental distinction or an opportunity for further deterministic improvements.

\paragraph{Our main techniques}
Our algorithms for 3- and 4-coloring combine reductions to constraint satisfaction problems (CSPs) with dynamic programming (DP) over tree and path decompositions, choosing the best running-time bound.
These tools are summarized in \cref{sec:CSP-and-DP}.
For 4-coloring, we also use a reduction to 3-coloring.
Before applying these approaches, we process high-degree structure by selecting a bushy forest for 3-coloring (\cref{sec:three-color-preprocess}) and vertex-disjoint $K_{1,3}$ packings for 4-coloring (\cref{sec:four-color-star-packing}), leaving residual graphs of bounded degree.

These approaches exploit complementary structural features.
CSP algorithms have running times exponential in the number of variables regardless of constraint density, making them useful for dense residual graphs.
Fixing colors on selected high-degree vertices further reduces the CSP cost by restricting the available colors of their neighbors.
In contrast, decomposition-based DP benefits from small treewidth or pathwidth, whose upper bounds depend on the degree structure of the residual graph and favor sparse instances.
Our analyses exploit this complementarity by identifying structural parameters that control the costs of the different approaches.

For 3-coloring, after selecting the bushy forest, we track
the total number of degree-three vertices and the
maximum number in any residual component.
A component with many degree-three vertices lets us
apply a stronger packing bound to more vertices,
improving the CSP bound.
Conversely, if every component contains few such vertices,
each has small pathwidth and can be solved efficiently
by DP after coloring the forest.
Thus, the maximum number per component captures
the trade-off between the CSP and DP bounds.

For 4-coloring, we strengthen the CSP approach using
a multiple-cover construction for triangles.
We construct nine list assignments such that every triangle
coloring is compatible with at least two of them
(\cref{lm:triangle-two-list-cover}), yielding
an amortized factor of $9/2$ per triangle instead of
the factor of $4.837$ used in
\cite{wu_et_al:LIPIcs.ESA.2024.103}.
We combine this improvement with structural parameters
tracking the numbers of residual cliques and the
distribution of star centers among them.
These parameters capture the trade-off between coloring
the selected stars and the resulting savings
in the CSP and decomposition-based DP algorithms.
Together with the reduction to 3-coloring, these approaches
provide complementary running-time bounds that yield
our improved bounds for 4-coloring.

\section{Preliminaries}
We use the notation $[x]_+=\max\{x,0\}$ and
write $\ln x$ for the natural logarithm of $x$.
The $O^*(\cdot)$ notation suppresses factors polynomial in $n$, where
$n$ denotes the number of vertices of the input graph.
Unless stated otherwise, all algorithms are deterministic.
Our randomized algorithms are one-sided Monte Carlo algorithms:
they always reject no-instances and accept yes-instances
with probability at least a fixed positive constant.
Independent repetition amplifies the success probability
to $1-2^{-\Omega(n)}$ with only polynomial overhead,
which is absorbed into the $O^*(\cdot)$ notation.
Whenever an algorithm makes multiple randomized calls,
we amplify each call sufficiently to bound the overall
failure probability by a union bound.

All graphs are undirected and simple; that is, they contain neither self-loops nor multiple edges.
For a graph $G=(V,E)$, we denote its vertex and edge sets by $V(G)$ and $E(G)$, respectively.
For a vertex $v\in V(G)$, we denote its degree by $\deg_G(v)$ and its neighborhood by $N_G(v)$.
We denote the maximum and minimum degrees of $G$ by
$\Delta(G)$ and $\delta(G)$, respectively.
For two vertices $u,v\in V(G)$, let $\operatorname{dist}_G(u,v)$
denote the minimum number of edges in a path between $u$ and $v$
in $G$, with $\operatorname{dist}_G(u,v)=\infty$ if no such path exists.
For $X\subseteq V(G)$, we denote by $G[X]$ the subgraph
of $G$ induced by $X$, that is,
$V(G[X])=X$ and
$E(G[X])=\{uv\in E(G)\mid u,v\in X\}$.
We write $G-X$ for $G[V(G)\setminus X]$ and define
$N_G(X)=\left(\bigcup_{v\in X}N_G(v)\right)\setminus X$.
For disjoint vertex sets $X_1,X_2\subseteq V(G)$, we define
$E_G(X_1,X_2)=\{uv\in E(G)\mid u\in X_1,\ v\in X_2\}$.

A \emph{tree decomposition} of a graph $G$ consists of a tree
$\mathcal{T}$ and a family of bags
$\{B_x\}_{x\in V(\mathcal{T})}$, where $B_x\subseteq V(G)$ for every
$x\in V(\mathcal{T})$ and
$\bigcup_{x\in V(\mathcal{T})} B_x=V(G)$, satisfying the following
conditions:
\begin{enumerate}[itemsep=1pt, parsep=0pt]
    \item For every $uv\in E(G)$, there exists a node
    $x\in V(\mathcal{T})$ such that $u,v\in B_x$.
    \item If $v\in B_i\cap B_j$, then $v\in B_x$ for every node $x$
    on the unique path in $\mathcal{T}$ between $i$ and $j$.
\end{enumerate}

The \emph{width} of a tree decomposition is the size of its largest
bag minus one.
A \emph{path decomposition} is a tree decomposition in which
$\mathcal{T}$ is a path, with the same notion of width.

For a nonnegative integer $k$, let $[k]=\{1,\ldots,k\}$,
with $[0]=\emptyset$.
A \emph{coloring} of a graph $G=(V,E)$ is a function
$\phi:V\to\mathbb{N}$ such that $\phi(u)\neq\phi(v)$ for every edge
$uv\in E$.
The \emph{$k$-coloring} problem asks whether $G$ admits a coloring
$\phi:V\to[k]$.
Throughout this paper, we consider the decision version of $k$-coloring.
The corresponding search version, which asks to output such a coloring
if one exists, can be solved within the same $O^*(\cdot)$ running time
by a standard self-reduction; see \cref{app:decision-to-search}.

A \emph{list $k$-coloring} instance consists of a graph $G$ and
a list $L(v)\subseteq[k]$ for each vertex $v\in V(G)$.
The \emph{list $k$-coloring} problem asks whether $G$ admits a coloring
$\phi$ such that $\phi(v)\in L(v)$ for every $v\in V(G)$.
List $k$-coloring arises naturally when extending a partial
$k$-coloring: each uncolored vertex is assigned the list of colors in
$[k]$ that are not used by its already colored neighbors.

For a positive integer $a$, we denote by $K_a$ the complete graph on
$a$ vertices.
For positive integers $a$ and $b$, we denote by $K_{a,b}$ the complete
bipartite graph with bipartition $(A,B)$, where $|A|=a$, $|B|=b$, and
$E(K_{a,b})=\{uv\mid u\in A,\ v\in B\}$.

To analyze branching algorithms, we use the measure-and-conquer method
\cite{fomin2009measure}.
A \emph{measure} is a nonnegative real-valued function
$\mu$ on instances.
Suppose that a branching rule applied to an instance $P$
produces $b$ subinstances $P_1,\ldots,P_b$ such that
$\mu(P)-\mu(P_i)\ge\beta_i>0$ for every $i\in[b]$.
The \emph{work factor} of the branching rule, denoted by
$\lambda(\beta_1,\ldots,\beta_b)$, is the unique positive real number
$x$ satisfying $\sum_{i=1}^b x^{-\beta_i}=1$.
If every branching rule has work factor at most $y>1$,
then $\sum_{i=1}^b y^{\mu(P_i)}\le y^{\mu(P)}$
for each branching rule.
Consequently, the overall running time is $O^*(y^{\mu(P)})$,
provided that each terminal instance $L$ can be solved in
$O^*(y^{\mu(L)})$ time.
For the 3-coloring preprocessing, we use the number of remaining vertices as the measure, whereas our 4-coloring analysis uses a weighted measure and directly applies the branching inequalities of \cite{wu_et_al:LIPIcs.ESA.2024.103}.

\section{CSP Algorithms and Decomposition-Based DP}
\label{sec:CSP-and-DP}

We first introduce the \emph{constraint satisfaction problem (CSP)}.

\begin{definition}[$(A,B)$-CSP]
An $(A,B)$-CSP instance consists of a set of variables, each with at most $A$ possible colors, and a set of constraints.
Each constraint specifies at most $B$ variable--color pairs and forbids the assignment in which every specified variable is assigned its corresponding color.
The task is to find an assignment of colors to the variables that satisfies all constraints.
\end{definition}

The $(k,2)$-CSP problem generalizes $k$-coloring.
For the algorithm of Beigel and Eppstein~\cite{BEIGEL2005168},
we define
\begin{equation}
\label{eq:CSP-Lambda}
\Lambda=\lambda(4,4,5,5)\approx1.36443.
\end{equation}

\begin{lemma}[{\cite[Theorem~1]{BEIGEL2005168}}]
\label{lm:3-2-CSP-time}
A $(3,2)$-CSP instance with $n$ variables can be solved
in $O^*(\Lambda^n)$ time.
\end{lemma}

We also use the following local reduction for variables with only two available colors.

\begin{lemma}[{\cite[Lemma~2]{BEIGEL2005168}}] \label{lm:3-2-CSP-local-reduction}
A variable with at most two available colors in an $(A,2)$-CSP instance can be eliminated in polynomial time while preserving satisfiability.
The reduction introduces constraints only among variables that shared a constraint with the eliminated variable, leaving all other original constraints unchanged.
\end{lemma}

Hence, by repeatedly applying this reduction, we may assume that every remaining variable in a $(3,2)$-CSP instance has three possible colors.

We next discuss dynamic programming (DP) over graph decompositions.
Given a tree decomposition of width $\mathrm{tw}$, $k$-coloring can be solved in $O^*(k^{\mathrm{tw}})$ time \cite{TelleProskurowski1997}.
The algorithm extends immediately to list $k$-coloring by forbidding disallowed colors in each bag state.

\begin{lemma}[\cite{TelleProskurowski1997}] \label{lm:treewidth-DP}
Given a graph $G$ and a tree decomposition of width $\mathrm{tw}$, list
$k$-coloring can be solved in $O^*(k^{\mathrm{tw}})$ time.
\end{lemma}

Jansen and Nederlof obtained the following randomized algorithm.
Although their theorem is stated for ordinary $k$-coloring, they
explicitly observe that it extends directly to list $k$-coloring by
assigning sufficiently large weights to forbidden vertex--color pairs;
see \cite[Section 3]{JANSEN2019520}.

\begin{lemma}[{\cite{JANSEN2019520}}] \label{lm:pathwidth-DP}
Given a graph $G$ and a path decomposition of width $\mathrm{pw}$,
there is a randomized algorithm that solves list $k$-coloring in
$O^*\left(
\left(\left\lfloor \Delta(G) / 2 \right\rfloor+1\right)^{\mathrm{pw}}
\right)$
time.
\end{lemma}

We use a polynomial-time algorithm to construct path decompositions of bounded width.
Although the bound in \cite[Lemma~1]{Fomin2009} is stated
for $n>n_\varepsilon$, where $n_\varepsilon$ depends only
on $\varepsilon$, it extends to all graphs with an
additive $O(1)$ term for every fixed $\varepsilon>0$,
since graphs with $n\le n_\varepsilon$ admit trivial
path decompositions of width at most $n_\varepsilon$.

\begin{restatable}[{\cite[Lemma~1]{Fomin2009}}]{lemma}{LmPathWidthUpperbound}
\label{lm:pathwidth-upperbound}
For every fixed $\varepsilon>0$ and every graph $G$ on $n$ vertices,
a path decomposition of $G$ of width at most
\[
\frac{1}{6}n_3
+\frac{1}{3}n_4
+\frac{13}{30}n_5
+\frac{23}{45}n_6
+n_{\ge 7}
+\varepsilon n+O(1)
\]
can be constructed in polynomial time.
Here, $n_i$ denotes the number of vertices of degree $i$ in $G$
for $i\in\{3,4,5,6\}$, and $n_{\ge 7}$ denotes the number of
vertices of degree at least $7$.
\end{restatable}

For a simpler analysis, we derive a coarser upper bound in
terms of the degree excess.

\begin{restatable}{lemma}{LmPathWidthDegreeExcessBound}
\label{lm:pathwidth-degree3-excess-bound}
For every fixed $\varepsilon>0$ and every graph $G$ on $n$ vertices, a path decomposition of $G$ of width at most
\[
\frac{1}{6} \sum_{v \in V(G)} [\deg_G(v) - 2]_+ + \varepsilon n + O(1)
\]
can be constructed in polynomial time.
\end{restatable}
\begin{proof}
See \cref{app:treewidth-degree3-excess-bound}.
\end{proof}

\section{Algorithms for 3-Coloring} \label{sec:three-coloring}

\subsection{Overview of the 3-Coloring Algorithm}

An overview of our 3-coloring algorithm is given in
\cref{fig:three-coloring-overview}.
In \cref{sec:three-color-preprocess}, we describe the
preprocessing and the construction of a maximal bushy forest.
In \cref{sec:three-color-CSP,sec:three-color-width-DP},
we derive running-time bounds for the CSP and width-based
DP algorithms, respectively.
Finally, in \cref{sec:three-color-LP}, we explain that,
since the logarithmic running-time bounds are linear
in the parameters, their worst-case minimum can be analyzed
using linear programming, and state the resulting upper bounds.
The explicit formulations and dual certificates are given
in \cref{app:three-coloring-LP}.

\begin{figure}[htbp]
\centering
\fbox{%
\begin{minipage}{0.98\linewidth}
\begin{enumerate}[itemsep=1pt, topsep=1pt, parsep=0pt]
\item Apply the preprocessing in
\cref{sec:three-color-preprocess} to obtain a maximal
bushy forest $F$.

\item Enumerate and fix the colors of the non-leaf vertices
of $F$.

\item To color the leaves of $F$ and the residual graph
$H=G-V(F)$, use the structural parameters to select the
algorithm with the smallest running-time bound among
the reduction to $(3,2)$-CSP
(\cref{lm:three-coloring-csp-packing}),
the tree decomposition DP
(\cref{lm:three-coloring-subcubic-width-DP,lm:three-coloring-treewidth-DP}),
and the randomized path decomposition DP
(\cref{lm:three-coloring-subcubic-width-DP}).
\end{enumerate}
\end{minipage}%
}
\caption{Overview of the 3-coloring algorithm.}
\label{fig:three-coloring-overview}
\end{figure}

\subsection{Preprocessing the Graph} \label{sec:three-color-preprocess}

We apply the preprocessing steps in \cref{fig:three-color-preprocessing}.
Steps~1--3 are also used in the preprocessing of the
$O^*(1.3289^n)$-time 3-coloring algorithm of
\cite{BEIGEL2005168}.
Step~1 is immediate, since any 3-coloring of $G-v$ extends
to $v$ when $\deg_G(v)\le2$.
Thus, we may assume $\delta(G)\ge3$.
Step~2 follows from \cref{lm:Beigel-Lemma21-degree3-cycle}.

\begin{lemma}[{\cite[Lemma~21]{BEIGEL2005168}}] \label{lm:Beigel-Lemma21-degree3-cycle}
Let $G$ be a graph containing a cycle whose vertices all have degree three.
Then the 3-coloring problem on $G$ can be reduced to smaller instances with work factor at most $\lambda(5,6,7,8)\le 1.2434$.
\end{lemma}

Since this work factor is smaller than the running-time bases in
\cref{thm:ThreeDet,thm:ThreeRand}, the branching does not increase
the running-time bounds.
Consequently, we may assume that the vertices of degree three
induce a forest.
This property is used to derive constraints in
\cref{lm:three-coloring-constraints}.

\begin{figure}[b]
\centering
\fbox{%
\begin{minipage}{0.98\linewidth}
\begin{enumerate}[itemsep=1pt, topsep=1pt, parsep=0pt]
\item Repeatedly remove vertices of degree at most two, so that $\delta(G)\ge 3$.

\item If the graph contains a cycle of degree-three vertices,
split the problem into smaller instances according to
\cref{lm:Beigel-Lemma21-degree3-cycle}, and recursively apply
Steps~1 and~2 to each resulting instance.

\item Greedily construct a maximal bushy forest $F$.

\item Modify $F$ in polynomial time so that it remains a maximal bushy forest and every tree containing a leaf adjacent to a vertex $v\notin V(F)$ with $\deg_{G-V(F)}(v)=3$ is a $K_{1,4}$.

\end{enumerate}
\end{minipage}%
}
\caption{Overview of the preprocessing steps.}
\label{fig:three-color-preprocessing}
\end{figure}

For Step~3, we extract a maximal bushy forest from $G$, following the idea introduced by Beigel and Eppstein in \cite[Section~6.2]{BEIGEL2005168}.
A \emph{bushy forest} is an unrooted forest contained in the
given graph in which every internal vertex, that is, every
non-leaf vertex, has degree at least four in the forest,
and every tree contains at least one internal vertex.
A bushy forest is \emph{maximal} if no internal vertex has a neighbor outside the forest, no leaf has three or more neighbors outside the forest, and no vertex outside the forest has four or more neighbors outside the forest.
A maximal bushy forest can be constructed greedily
by repeatedly applying the following operations: if an internal vertex has a neighbor outside the forest, we add that neighbor to its tree as a new leaf; if a leaf has at least three neighbors outside the forest, we add all of these neighbors to its tree; and if a vertex outside the forest has at least four neighbors outside the forest, we add that vertex and all of its neighbors outside the forest as a new tree. This construction runs 
in polynomial time.

For Step~4, we use a weaker variant of the preprocessing described in the preprint \cite[Lemma~3.1]{meijer20233coloringtimeo13217n} and provide a self-contained proof of this variant in \cref{app:lemma-bushy-forest-normalize}.

\begin{restatable}[{\cite{meijer20233coloringtimeo13217n}}]{lemma}{LmBushyForestNormalize} \label{lm:bushy-forest-normalize}
Given a maximal bushy forest $F$, we can modify $F$ in polynomial time so that it remains a maximal bushy forest and every tree containing a leaf adjacent to a vertex $v\notin V(F)$ with $\deg_{G-V(F)}(v)=3$ is a $K_{1,4}$.    
\end{restatable}

\begin{figure}[t]
    \centering
    \includegraphics[width=0.9\linewidth]{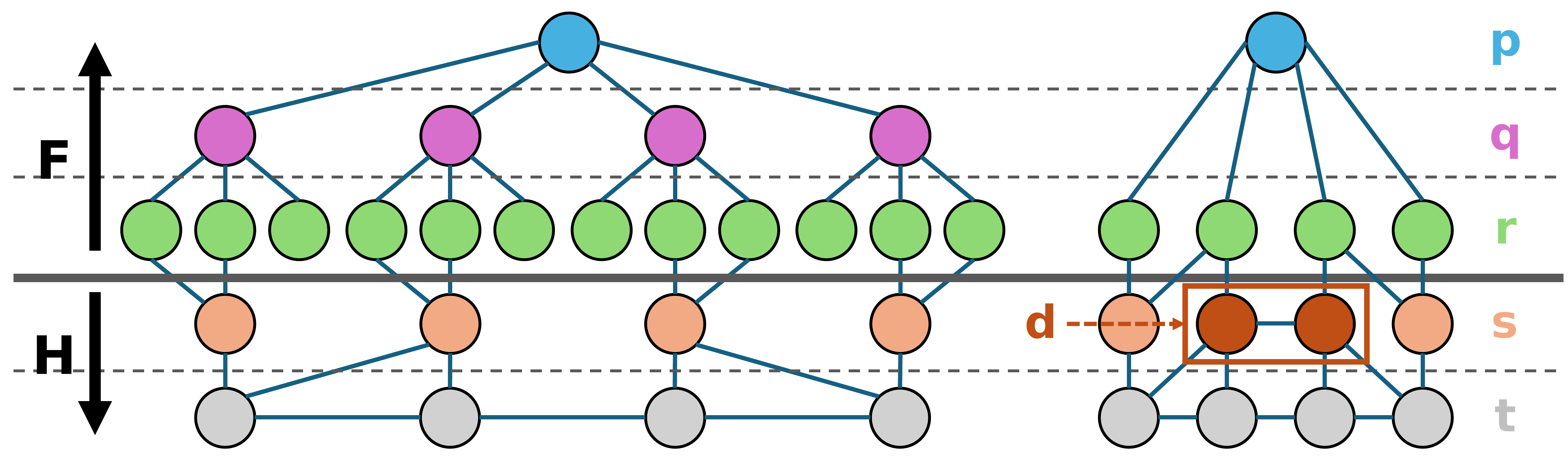}
    \caption{Illustration of the parameters $p,q,r,s,t,d$ associated with a maximal bushy forest $F$.}
    \label{fig:three-color-vertices-partition}
\end{figure}

We partition the vertices of $G$ according to a maximal bushy forest $F$.
For each tree of $F$, we arbitrarily choose an internal vertex as its root.
Let $p$, $q$, and $r$ denote the numbers of roots, non-root internal vertices, and leaves, respectively.
We denote the set of leaves by $R$, so that $|R|=r$.
Let $H=G-V(F)$ be the residual graph, and define $S=N_G(R)\cap V(H)$ and $T=V(H)\setminus S$, with $s=|S|$ and $t=|T|$.
The parameters $p,q,r,s,t$ follow the notation of \cite[Section~6.5]{BEIGEL2005168}.
We further define $D=\{v\in S\mid \deg_H(v)=3\}$ and $d=|D|$.
The parameters are illustrated in \cref{fig:three-color-vertices-partition}.

\begin{lemma} \label{lm:H-degree-3}
The graph $H$ is subcubic, that is, $\Delta(H)\le3$, and its degree-three vertices are exactly the vertices in $D\cup T$.
In particular, $H$ has exactly $d+t$ vertices of degree three.
Moreover, every vertex $v\in T$ satisfies $\deg_H(v)=\deg_G(v)=3$.
\end{lemma}

\begin{proof}
By the maximality of the bushy forest $F$, every vertex $v\in V(H)$ satisfies $\deg_H(v)\le3$.
Moreover, a vertex $v\in T$ is not adjacent to any leaf of $F$ by the definition of $T$, and it is not adjacent to any internal vertex of $F$ by the maximality of $F$.
Hence, $v$ has no neighbor in $F$, and therefore $\deg_H(v)=\deg_G(v)$.
By Step~1 of \cref{fig:three-color-preprocessing}, we have $\delta(G)\ge3$, and thus $\deg_H(v)=\deg_G(v)=3$ for every $v\in T$.
By the definition of $D$, the degree-three vertices in $S$ are exactly the vertices in $D$.
Therefore, the degree-three vertices of $H$ are exactly those in $D\cup T$, and their number is $d+t$.
\end{proof}

The parameters defined above satisfy the following constraints.

\begin{lemma}
\label{lm:three-coloring-constraints}
The parameters $p,q,r,s,t,d$ are nonnegative and satisfy
\begin{equation}
\label{eq:three-coloring-constraints}
p+q+r+s+t=n,\quad d\le s,\quad 4p+2q\le r,\quad
s\le2r,\quad d\le8p,\quad t\le2s+d.
\end{equation}
\end{lemma}

\begin{proof}
All six parameters are cardinalities and hence nonnegative.
The sets counted by $p,q,r$ partition $V(F)$, while
$S$ and $T$ partition $V(H)$, giving
$p+q+r+s+t=n$. $D \subseteq S$ gives $d \le s$.

By the definition of a bushy forest, each tree has
at least four leaves, and each additional internal
vertex contributes at least two more leaves.
Thus, $4p+2q\le r$.
Moreover, by maximality, each leaf has at most two
neighbors outside $F$, giving $s\le2r$.

To show $d\le8p$, let $R_D=N_G(D)\cap R$.
By Step~4 of \cref{fig:three-color-preprocessing},
every tree containing a vertex of $R_D$ is a $K_{1,4}$.
Hence $|R_D|\le4p$.
Since every vertex in $D$ is adjacent to a vertex
of $R_D$, and each vertex of $R_D$ has at most two
neighbors outside $F$, we obtain
$d\le2|R_D|\le8p$.

Finally, by Step~2 of \cref{fig:three-color-preprocessing}
and \cref{lm:H-degree-3}, $G[T]$ is a forest,
so $|E(G[T])|\le t$.
Since every vertex in $T$ has degree three in $H$,
we have $|E_G(T,S)|=3t-2|E(G[T])|\ge t$.
On the other hand, vertices in $S\setminus D$
have degree at most two in $H$, while those in $D$
have degree three, giving
$|E_G(S,T)|\le2(s-d)+3d=2s+d$.
Therefore, $t\le2s+d$.
\end{proof}

We first enumerate the colorings of the internal vertices of $F$.
Each root has at most three choices, and each non-root
internal vertex has at most two choices once its parent
is colored.
Thus, the number of such colorings is at most
\begin{equation}
\label{eq:internal-colorings}
B_I=3^p2^q.
\end{equation}
For each such coloring, every leaf in $R$ has at most
two available colors.
In the following sections, we focus on coloring the
remaining leaves $R$ and the residual graph $H$.

\subsection{Algorithms via CSP} \label{sec:three-color-CSP}
After coloring the internal vertices of $F$,
it remains to color the leaves $R$ and the residual graph $H$.
A simple approach is to formulate the remaining problem
as a $(3,2)$-CSP instance.
Since each leaf has at most two available colors,
the corresponding variables can be eliminated in polynomial time
by \cref{lm:3-2-CSP-local-reduction},
leaving at most $|V(H)|=s+t$ three-valued variables.
By \cref{lm:3-2-CSP-time}, the resulting instance can be solved
in $O^*(\Lambda^{s+t})$ time, yielding a total running time of
$O^*(B_I\Lambda^{s+t})$.

To improve this bound, we process $K_{1,3}$ stars
centered at degree-three vertices of $H$.
For each star, branching on the color of its center
requires at most three choices and reduces the domains
of its three leaves to size at most two.
By \cref{lm:3-2-CSP-local-reduction}, these leaves
can then be eliminated.
Thus, all four vertices of the $K_{1,3}$ are processed
at a cost of at most three choices, compared with
$\Lambda^4$ when applying $(3,2)$-CSP directly.
The relative cost is therefore $3\Lambda^{-4}<1$.

A $k$-\emph{packing} in a graph $G$ is a set
$P\subseteq V(G)$ such that
$\operatorname{dist}_G(u,v)\ge k+1$
for all distinct $u,v\in P$.
To obtain vertex-disjoint copies of $K_{1,3}$,
we seek a 2-packing consisting of degree-three vertices of $H$.
Since $\Delta(H)\le3$, each vertex has at most
$1+3+6=10$ vertices within distance two.
Thus, a greedy algorithm finds a 2-packing containing
at least $(d+t)/10$ degree-three vertices of $H$.
A stronger bound for connected cubic graphs was obtained
using a refined greedy algorithm~\cite{GODDARD2024100850}.

\begin{lemma}[\cite{GODDARD2024100850}]
\label{lm:cubic-packing}
For every connected cubic graph $G$ on $n$ vertices,
a 2-packing of size at least
$\frac{17}{132}(n-3)$
can be constructed in polynomial time.
\end{lemma}

We extend their 2-packing algorithm to connected subcubic graphs.

\begin{lemma}
\label{lm:subcubic-packing}
Suppose that, for some constant $\alpha>0$, every
connected cubic graph on $n$ vertices admits a 2-packing
of size at least $\alpha n-O(1)$ that can be constructed
in polynomial time.
Then every connected subcubic graph $G$ admits a 2-packing
of size at least $\alpha n_3-O(1)$ consisting entirely
of degree-three vertices of $G$, where $n_3$ denotes
the number of such vertices.
This packing can be constructed in polynomial time.
\end{lemma}
\begin{proof}

If $n_3=0$, the empty packing suffices.
Hence, assume $n_3>0$.
We construct an auxiliary graph $J$ on the degree-three
vertices of $G$, joining two distinct vertices
if a path in $G$ connects them using only degree-two
internal vertices.
Since $G$ is connected, so is $J$.
Moreover,
$\operatorname{dist}_J(u,v)\le\operatorname{dist}_G(u,v)$
for all $u,v\in V(J)$.
Thus, every 2-packing in $J$ is also a 2-packing in $G$.

Since $J$ is subcubic but not necessarily cubic,
we augment it to a cubic graph as follows.
We take two copies of the current graph and join each
pair of corresponding vertices whose degrees are less
than three by an edge.
This increases the degree of each vertex from $d$
to $\min\{3,d+1\}$.
We repeat this operation until the graph becomes cubic.
Since at most three iterations are needed, the resulting
cubic graph $J'$ contains $n_{\mathrm{cp}}\le 8$
copies of $J$.
By assumption, we can construct a 2-packing $P$ in $J'$
of size at least $\alpha |V(J')|-O(1)$.
By the pigeonhole principle, some copy of $J$
contains at least $|P|/n_{\mathrm{cp}}$ vertices of $P$.
These vertices form a 2-packing in $J$, since
distances within a copy of $J$ cannot be smaller
than those in $J'$.
Moreover, every 2-packing in $J$ is also a 2-packing in $G$.
Since $|V(J')|=n_{\mathrm{cp}}|V(J)|$, we obtain
a 2-packing in $G$ consisting of degree-three vertices
of size at least
$\frac{\alpha |V(J')|-O(1)}{n_{\mathrm{cp}}}
=\alpha |V(J)|-O(1)
=\alpha n_3-O(1)$.
\end{proof}

Thus, components containing many degree-three vertices
allow us to exploit the stronger packing bound, whereas
those containing few degree-three vertices yield smaller
pathwidth in the width-based algorithm described later.
To exploit this trade-off, let $c$ denote the maximum
number of degree-three vertices in any component of $H$.
By definition, we have
\begin{equation}
\label{eq:three-color-connected-component}
0\le c\le d+t.
\end{equation}
We apply \cref{lm:cubic-packing,lm:subcubic-packing}
to a component of $H$ containing the most degree-three
vertices, 
and use the simple greedy algorithm with a $1/10$
guarantee for the others.
Combining these 2-packings yields a 2-packing of degree-three
vertices of size at least
\[
\frac{17}{132}c-O(1)+\frac{d+t-c}{10}
=\frac{d+t}{10}+\frac{19}{660}c-O(1).
\]
Applying the branching with relative cost $3\Lambda^{-4}$
to each 2-packing vertex yields the following bound.
\begin{lemma}
\label{lm:three-coloring-csp-packing}
For a graph $G$ after preprocessing, 3-coloring
can be solved in time
\[
O^*\left(
B_I\Lambda^{s+t}
(3\Lambda^{-4})^{\frac{d+t}{10}+\frac{19}{660}c}
\right).
\]
\end{lemma}

\subsection{Algorithms via Decomposition-Based DP} \label{sec:three-color-width-DP}

We take a maximal matching $M$ of $G[R]$ and let $m=|M|$. By definition, we have 
\begin{equation} \label{eq:three-color-matching-constraint}
0 \le m \le r/2.
\end{equation}
After coloring the internal vertices of $F$, each leaf has at most two available colors. Since these colors belong to the common color set $[3]$, the endpoints of each edge in $M$ admit at most three colorings. Thus, the endpoints of all edges in $M$ can be colored in at most $3^m$ ways.

Let $U=R\setminus V(M)$ be the set of unmatched leaves. After coloring the internal vertices of $F$ and the endpoints of $M$, it remains to color $U$ and $H$. We consider two approaches: coloring $U$ first or processing it together with $H$.

\paragraph{Coloring $U$ first.}
We first color $U$ in at most $2^{|U|}=2^{r-2m}$ ways,
since each leaf has at most two available colors.
For each coloring of $F$, we assign to each vertex
$v\in V(H)$ the list of colors in $[3]$
not used by its colored neighbors.
We then solve list 3-coloring on each connected
component of $H$ independently.
Recall that $c$ is the maximum number of degree-three
vertices in a component of $H$, and that
$\Delta(H)\le 3$ by \cref{lm:H-degree-3}.
By \cref{lm:pathwidth-upperbound}, each component admits
a path decomposition of width at most
$c/6+\varepsilon n+O(1)$.
Thus, all components can be solved in deterministic time
$O^*(3^{c/6+\varepsilon n})$
by \cref{lm:treewidth-DP}, or in randomized time
$O^*(2^{c/6+\varepsilon n})$
by \cref{lm:pathwidth-DP}.

\begin{lemma}
\label{lm:three-coloring-subcubic-width-DP}
Let $B_R=3^m2^{r-2m}$ be an upper bound on the number
of colorings of $R$ for each coloring of the internal
vertices of $F$.
For every fixed $\varepsilon>0$, a graph $G$ after
preprocessing can be 3-colored in deterministic time
$O^*(B_IB_R\,3^{c/6+\varepsilon n})$
or randomized time
$O^*(B_IB_R\,2^{c/6+\varepsilon n})$.
\end{lemma}

\paragraph{Processing $U$ together with $H$.}
We consider coloring $K=G[U\cup V(H)]$.
Since $K$ need not be subcubic,
we can no longer exploit the subcubic bound
in \cref{lm:pathwidth-DP},
but \cref{lm:treewidth-DP} remains useful.
Since distinct components of $H$ may be connected through $U$,
we use $d+t$ rather than $c$ to bound the width
of a tree decomposition of $K$ using
\cref{lm:pathwidth-degree3-excess-bound}.
By the maximality of $F$, each vertex $v\in U\subseteq R$
has at most two neighbors outside $F$.
The maximality of $M$ also implies that $U$ is independent.
Hence, $\deg_K(v)\le2$ for every $v\in U$.
Thus, vertices in $U$ contribute no degree excess
above two, and each edge between $U$ and $H$
increases this excess by at most one. Hence,
\[
\sum_{v\in V(K)}[\deg_K(v)-2]_+
\le |D\cup T| + |E_G(U,V(H))|
\le d+t+2|U|=d+t+2r-4m.
\]
Therefore, the width of the tree decomposition of $K$
is at most $(d+t+2r-4m)/6+\varepsilon n+O(1)$.
Combining the $3^m$ colorings of the endpoints of $M$
with the list 3-coloring algorithm of
\cref{lm:treewidth-DP} on $K$ yields the following bound.
\begin{lemma}
\label{lm:three-coloring-treewidth-DP}
For every fixed $\varepsilon>0$, 3-coloring on a graph
$G$ after preprocessing can be solved in time
\[
O^*\left(
B_I3^m\,3^{(d+t+2r-4m)/6+\varepsilon n}
\right).
\]
\end{lemma}

\subsection{Running-Time Analysis via Linear Programming} \label{sec:three-color-LP}
We use the fastest of the algorithms in
\cref{lm:three-coloring-csp-packing,lm:three-coloring-subcubic-width-DP,lm:three-coloring-treewidth-DP}.
The logarithm of each running-time bound, ignoring
polynomial factors, is linear in the nonnegative
parameters $p,q,r,s,t,d,c,m$.
Normalizing these parameters through division by $n$,
we formulate a linear program based on the constraints in
\eqref{eq:three-coloring-constraints},
\eqref{eq:three-color-connected-component}, and
\eqref{eq:three-color-matching-constraint}
to determine $\tau$ in the worst-case
running-time bound $O^*(e^{\tau n})$.

Since only the $O^*(B_IB_R2^{c/6+\varepsilon n})$ bound in
\cref{lm:three-coloring-subcubic-width-DP} is randomized,
we exclude it from the deterministic analysis
and include it in the randomized analysis.
We fix $\varepsilon=10^{-9}$ and let
$\tau_{\mathrm{det}}$ and $\tau_{\mathrm{rand}}$
denote the optimal values of the deterministic and
randomized linear programs, respectively.
The detailed LP formulations and dual certificates for the
resulting bounds are given in \cref{app:three-coloring-LP}.
Here we state the results:
$\tau_{\mathrm{det}}\le\ln\ThreeDetBase$ and
$\tau_{\mathrm{rand}}\le\ln\ThreeRandBase$,
which yield the following theorems.

\ThmThreeDet*
\ThmThreeRand*

\section{Algorithms for 4-Coloring}

\subsection{Overview of the 4-Coloring Algorithm}

An overview of our 4-coloring algorithm is given in \cref{fig:four-color-overview}.
Following the idea of \cite{wu_et_al:LIPIcs.ESA.2024.103},
we consider an instance $P=(G,I,S)$, where $I$ and $S$ are
disjoint subsets of $V(G)$, $I$ is an independent set, and
$N_G(I)\subseteq S$.
The vertices in $I$ are assigned color $4$, those in $S$
use colors from $[3]$, and those in
$U=V(G)\setminus(I\cup S)$ may use any color in $[4]$.
Thus, it suffices to color $G[S\cup U]$ subject to
these color restrictions.
An instance is \emph{terminal} if every component
of $G[U]$ is $K_3$ or $K_4$.
Starting from $(G,\emptyset,\emptyset)$, the branching
reductions of \cite{wu_et_al:LIPIcs.ESA.2024.103} replace nonterminal instances with smaller
instances until all resulting instances are terminal.
We apply their branching reductions unchanged.

\begin{figure}[t]
\centering
\fbox{%
\begin{minipage}{0.98\linewidth}
\begin{enumerate}[itemsep=1pt, topsep=1pt, parsep=0pt]
    \item Apply the branching reductions of
    \cite{wu_et_al:LIPIcs.ESA.2024.103} to reduce the initial
    instance $(G,\emptyset,\emptyset)$ to terminal instances.

    \item For each terminal instance, construct the two
    $K_{1,3}$-packings described in
    \cref{sec:four-color-star-packing}.

    \item Based on the structural parameters of each terminal
    instance, select the algorithm with the smallest
    running-time bound among the reduction to 3-coloring
    (\cref{lm:four-color-via-three-color}),
    the $(3,2)$-CSP algorithm
    (\cref{lm:four-color-via-csp}),
    the tree decomposition DP
    (\cref{lm:four-color-via-tree-width}),
    and the randomized path decomposition DP
    (\cref{lm:four-color-via-path-width}).
    Accept if any terminal instance admits a coloring.
\end{enumerate}
\end{minipage}%
}
\caption{Overview of the 4-coloring algorithm.}
\label{fig:four-color-overview}
\end{figure}

Our main technical contribution for 4-coloring is to improve
the running time for solving terminal instances.
In \cref{sec:four-color-terminal-to-overall}, we describe
the measure-and-conquer framework of
\cite{wu_et_al:LIPIcs.ESA.2024.103} and the convex optimization
used to translate terminal running-time bounds into an
overall bound for 4-coloring.
For an instance $P=(G,I,S)$, let $n_{k3}$ and $n_{k4}$ denote the numbers of $K_3$-components and $K_4$-components of $G[U]$, respectively.

\subsection{Preprocessing the Terminal Instances} \label{sec:four-color-star-packing}

For a graph $H$, an $H$-\emph{packing} in $G$ is a family of
pairwise vertex-disjoint subgraphs of $G$, each isomorphic to $H$.
We use maximal $K_{1,3}$-packings as a common preprocessing
step.

\paragraph{Two maximal $K_{1,3}$-packings}
We first greedily construct a maximal family of vertex-disjoint
copies of $K_{1,3}$ in $G[S]$.
Let $S_{\star 1}$ be the union of their vertex sets.
We then construct another maximal family of vertex-disjoint copies of
$K_{1,3}$ with centers in $U$ and leaves in
$S\setminus S_{\star 1}$.
Let $U_{\star 2}$ and $S_{\star 2}$ be its sets of centers
and leaves, respectively, and set
$S'=S\setminus(S_{\star 1}\cup S_{\star 2})$.
Write $s_{\star 1}=|S_{\star 1}|$,
$u_{\star 2}=|U_{\star 2}|$, and $s'=|S'|$, so that
$|S_{\star 2}|=3u_{\star 2}$ and
$|S|=s_{\star 1}+3u_{\star 2}+s'$.
The maximality of the packings implies the following.

\begin{lemma} \label{lm:four-color-star-maximality}
We have $\Delta(G[S\setminus S_{\star 1}])\le 2$ and
$|N_G(u)\cap S'|\le 2$ for every $u\in U\setminus U_{\star 2}$.
\end{lemma}

\begin{proof}
Otherwise, a vertex with three neighbors in
$S\setminus S_{\star 1}$ could form an additional
$K_{1,3}$, contradicting maximality of the first $K_{1,3}$-packing.
Similarly, a vertex $u\in U\setminus U_{\star 2}$
with three neighbors in $S'$ would extend the second $K_{1,3}$-packing.
\end{proof}

\paragraph{Classification of clique components}
We classify the clique components of $G[U]$ according to
the number of their vertices in $U_{\star 2}$.
Let $a_i$ denote the number of $K_3$-components containing
exactly $i$ vertices of $U_{\star 2}$ for $i=0,1,2,3$.
Similarly, let $b_i$ denote the number of such $K_4$-components
for $i=0,1,2,3,4$.
Then
\begin{equation} \label{eq:four-color-parameters}
n_{k3} = \sum_{i=0}^3 a_i,\quad
n_{k4} = \sum_{i=0}^4 b_i,\quad
u_{\star 2} = \sum_{i=1}^3 ia_i+\sum_{i=1}^4 ib_i,\quad
|S| = s_{\star 1}+3u_{\star 2}+s'.
\end{equation}

\subsection{Terminal Algorithms via 3-Coloring}

We give a simple reduction to 3-coloring.

\begin{lemma} \label{lm:four-color-via-three-color}
Suppose that 3-coloring on $n$-vertex graphs can be solved in
$O^*(\tau_3^n)$ time.
Then a terminal instance of 4-coloring can be solved in time
$
O^*\!\left(
3^{n_{k3}}4^{n_{k4}}
\tau_3^{|S|+2n_{k3}+3n_{k4}}
\right)
$.
\end{lemma}
\begin{proof}
Recall that the vertices of $I$ are assigned color $4$, whereas
those of $S$ are restricted to colors in $[3]$.
Since $N_G(I)\subseteq S$, there are no edges between $I$ and $U$.
We may therefore assume that each of the disjoint clique components
of $G[U]$ contains exactly one vertex of color $4$: if a component
does not use color $4$, we may recolor any one of its vertices with $4$.
We enumerate the vertex assigned color $4$ in each component of $G[U]$.
Since $G[U]$ consists of $n_{k3}$ copies of $K_3$ and $n_{k4}$
copies of $K_4$, there are $3^{n_{k3}}4^{n_{k4}}$ choices.
For each choice, the remaining vertices of $U$ and all vertices
of $S$ must be colored from $[3]$.
Thus, we can apply the assumed 3-coloring algorithm to the graph
induced by these $|S|+2n_{k3}+3n_{k4}$ vertices,
yielding the claimed running time.
\end{proof}

\subsection{Terminal Algorithms via CSP}

To reduce a terminal instance to $(3,2)$-CSP, we restrict
each vertex in $U$ to at most three colors.
We branch into list assignments while preserving satisfiability.
For each triangle $C$ with $V(C)\cap U_{\star 2}=\emptyset$,
we restrict its vertices to two-element lists.
Each list belongs to
$\binom{[4]}{2}
=\{\{1,2\},\{1,3\},\{1,4\},$
$\{2,3\},\{2,4\},\{3,4\}\}$.
For comparison, processing $t$ triangles incurs a factor
of $4.837^t$ in \cite{wu_et_al:LIPIcs.ESA.2024.103},
whereas our construction reduces this factor to
$(9/2)^t e^{o(n)}$.

\begin{lemma}
\label{lm:triangle-two-list-cover}
Let $C_1,\dots,C_n$ be pairwise vertex-disjoint triangles, and let
$V=\bigcup_{i=1}^n V(C_i)$.
There is a deterministic algorithm that constructs a family
$\mathcal L_n$ of list assignments $L:V\to\binom{[4]}{2}$
such that every coloring $\varphi:V\to[4]$ satisfying
$4\in\varphi(V(C_i))$ for all $i\in[n]$
is compatible with some $L\in\mathcal L_n$, meaning that
$\varphi(v)\in L(v)$ for every $v\in V$.
Moreover, $|\mathcal L_n|\le (9/2)^n e^{o(n)}$
and the family can be enumerated in
$O^*(|\mathcal L_n|)$ time.
\end{lemma}

\begin{proof}
As in the proof of \cref{lm:four-color-via-three-color}, we consider only colorings that use color $4$ on every triangle.
For a single triangle $C$, each $c\in[3]$ and each
$x\in V(C)$ define a list assignment
\[
L_{c,x}(v)=
\begin{cases}
[3]\setminus\{c\}, & v=x,\\
\{c,4\}, & v\neq x.
\end{cases}
\]
There are nine such list assignments.
Every coloring of $C$ using color $4$ is compatible with exactly two.
Indeed, let $v_4,v_a,v_b$ receive colors $4,a,b$, respectively,
and let $\{d\}=[3]\setminus\{a,b\}$.
These two assignments are
\[
\begin{array}{c|ccc}
 & v_4 & v_a & v_b\\ \hline
L_{a,v_b} & \{4,a\} & \{4,a\} & \{b,d\}\\
L_{b,v_a} & \{4,b\} & \{a,d\} & \{4,b\}
\end{array}
\]
There are no others: $x$ cannot be $v_4$, and either remaining
choice of $x$ uniquely determines $c$.

For $n$ triangles, the local choices give $9^n$ list assignments,
each coloring being compatible with exactly $2^n$.
Thus, a uniform sample preserves any fixed coloring with
probability $(2/9)^n$, and $O((9/2)^n)$ independent samples
preserve it with constant probability.

We derandomize this step by constructing $(9/2)^n e^{o(n)}$
list assignments such that every coloring is compatible with
at least one of them.
Consider a block of $j$ triangles.
Taking products of the nine local choices gives $9^j$
candidate list assignments.
The block has $18^j$ colorings, each compatible with
exactly $2^j$ candidates.

We select candidates greedily. Suppose $M$ colorings are not
yet compatible with any selected candidate. Each is compatible with exactly $2^j$ of the $9^j$ candidates.
By the pigeonhole principle, some candidate is compatible with
at least $M(2/9)^j$ of these colorings. Selecting it leaves at most
$M(1-(2/9)^j)$ colorings. Initially there are $18^j$ colorings. Thus, after $h$ selections, at most
\[
18^j\bigl(1-(2/9)^j\bigr)^h
< 18^j\exp\bigl(-h(2/9)^j\bigr)
\]
colorings remain.
Consequently, every coloring is compatible with a selected
candidate after at most $B_j$ selections, where
$
B_j=
\left\lceil
\left(\frac92\right)^j j\ln 18
\right\rceil
$.

Set $m=\lceil\ln (n+2)\rceil$ and partition the $n$ triangles
into $b=\lceil n/m\rceil$ blocks of sizes
$j_1,\dots,j_b$, where $1\le j_\ell\le m$.
For each block, select at most $B_{j_\ell}$ candidates as above,
and let $\mathcal L_n$ be the Cartesian product of the selected
families. Every coloring is compatible with some member of
$\mathcal L_n$, because its restriction to each block is
compatible with a selected candidate. Moreover,
\[
|\mathcal L_n|
\le \prod_{\ell=1}^b B_{j_\ell}
\le \left(\frac92\right)^n(m\ln 18+1)^b
= \left(\frac92\right)^n
\exp\left(
O\left(\frac{n\ln\ln n}{\ln n}\right)
\right)
= \left(\frac92\right)^n e^{o(n)}.
\]

Since $m=O(\log n)$, the colorings and candidates of each block
can be enumerated in polynomial time. We can therefore make
each greedy selection in polynomial time and enumerate
$\mathcal L_n$ in polynomial time per member, giving the
claimed enumeration time.
\end{proof}

We are now ready to bound the running time.

\begin{lemma}
\label{lm:four-color-via-csp}
There is an
$O^*\!\left(
3^{s_{\star 1}/4}
(\prod_{i=0}^{3}g_{3,i}^{a_i})
(\prod_{i=0}^{4}g_{4,i}^{b_i})
\Lambda^{s'}
e^{o(n)}
\right)$-time algorithm for solving a terminal instance of
4-coloring, where, for $m\in\{3,4\}$,
\[
g_{m,i}=
\begin{cases}
3m/2, & i=0,\\[2pt]
\Lambda^{m-1}(3+\Lambda^3), & i=1,\\[2pt]
\displaystyle
\frac{6}{(4-i)!}(4-i+i\Lambda^3), & 2\le i\le m.
\end{cases}
\]
\end{lemma}
\begin{proof}
For each clique $C$ of $G[U]$, we form a block consisting
of $V(C)$ and the leaves of each second star
centered in $V(C)\cap U_{\star 2}$.
We partition $S\cup U$ into the first stars
$S_{\star 1}$, the cliques with their associated
star leaves, and the remaining vertices $S'$.
We process these parts as follows.

For each first star, we choose one of the three colors in
$[3]$ for its center, since $S_{\star 1}\subseteq S$.
Its three leaves then have lists of size at most two and
can be eliminated by \cref{lm:3-2-CSP-local-reduction}.
As there are $s_{\star 1}/4$ first stars, this contributes
a factor of $3^{s_{\star 1}/4}$ to the overall running time.

For each clique $C$ of $G[U]$, let $m=|V(C)|$ and
$i=|V(C)\cap U_{\star 2}|$.
We show below that processing $C$ together with its
associated star leaves incurs a total branch cost
of at most $g_{m,i}$, where each branch has cost
$\Lambda^t$ and $t$ is the number of three-valued
variables left by this part.
By the definitions of $a_i$ and $b_i$, the cliques
together with their associated star leaves contribute
a factor of at most
$
\left(\prod_{i=0}^{3}g_{3,i}^{a_i}\right)
\left(\prod_{i=0}^{4}g_{4,i}^{b_i}\right)e^{o(n)}
$
to the overall running time.
The additional $e^{o(n)}$ factor comes from the
size bound in \cref{lm:triangle-two-list-cover}.

We leave the vertices of $S'$ unchanged, without branching
on their colors or restricting their lists.
Since $S'\subseteq S$, each has at most three available colors.
Thus, they contribute a factor of at most $\Lambda^{s'}$
to the overall running time. Multiplying the factors for the three types of vertices gives the claimed running-time bound.

\paragraph{Proof of the bounds on $g_{m,i}$.}
First consider $i=0$.
For $m=3$, \cref{lm:triangle-two-list-cover} gives
$9/2$ list assignments per triangle, up to an $e^{o(n)}$
factor. Every vertex then has a two-element list, so the
triangle adds no three-valued variables to the remaining
$(3,2)$-CSP instance. Hence $g_{3,0}=9/2$.
For $m=4$, choose two vertices of the $K_4$ to receive
the list $\{1,2\}$ and give the other two the list $\{3,4\}$.
There are $\binom{4}{2}=6$ choices, and every coloring of
the $K_4$ is compatible with exactly one of them.
No three-valued variables remain in the clique, so
$g_{4,0}=6$. The same six-way restriction of $K_4$ is used in
\cite[Section 3.1]{wu_et_al:LIPIcs.ESA.2024.103}.

Next, suppose $i=1$, and let $\{v\}=V(C)\cap U_{\star 2}$.
We fix the color of $v$.
If $v$ receives color $4$, each of its three second-star
leaves retains a three-element list, while the other $m-1$
vertices of $C$ have lists of size at most three.
This branch costs at most $\Lambda^{3+m-1}$.
If $v$ receives a color in $[3]$, its leaves have lists of
size at most two. There are three such choices, each costing
at most $\Lambda^{m-1}$ for the remaining vertices of $C$.
Hence
$
g_{m,1}
=\Lambda^{3+m-1}+3\Lambda^{m-1}
=\Lambda^{m-1}(3+\Lambda^3).
$

Finally, suppose $i\ge2$.
We fix the colors of all $i$ vertices in
$V(C)\cap U_{\star 2}$, assigning them distinct colors.
If none receives color $4$, there are
$\frac{3!}{(3-i)!}$ assignments when $i\le3$, and none when $i=4$.
In either case, the number of assignments is
$\frac{6(4-i)}{(4-i)!}$. The leaves of these centers then have lists
of size at most two, so each assignment costs $1$.
If one center receives color $4$, there are
$
i\frac{3!}{(4-i)!}
=\frac{6i}{(4-i)!}
$
assignments. Its three leaves retain three colors, while
the other centers' leaves have lists of size at most two.
Each assignment therefore costs $\Lambda^3$.
In either case, the remaining $m-i\le2$ vertices of $C$
have lists of size at most two. Thus,
$
g_{m,i}
=
\frac{6(4-i)}{(4-i)!}
+\frac{6i}{(4-i)!}\Lambda^3
=
\frac{6}{(4-i)!}(4-i+i\Lambda^3)
$.
\end{proof}

\subsection{Terminal Algorithms via Tree Decomposition DP}

We enumerate the colors of all vertices in $S_{\star 1}\cup U$.
Each first star has at most $3\cdot 2^3=24$ colorings.
Each $K_3$-component of $G[U]$ has $18$ colorings that
use color $4$, while each $K_4$-component has $24$ colorings.
Thus, we enumerate at most $24^{s_{\star 1}/4}18^{n_{k3}}24^{n_{k4}}$
colorings, discarding any that violate an edge between
precolored vertices.
For each remaining choice, we obtain a list 3-coloring instance
on $G[S\setminus S_{\star 1}]$ by giving each vertex the list of
colors in $[3]$ not used by its precolored neighbors.
By \cref{lm:four-color-star-maximality},
$\Delta(G[S\setminus S_{\star 1}])\le 2$.
Hence this graph is a disjoint union of paths, cycles, and
isolated vertices, and has treewidth at most two.
The list instance can therefore be solved in polynomial time
by \cref{lm:treewidth-DP}.
Therefore, we obtain the following bound.

\begin{lemma} \label{lm:four-color-via-tree-width}
A terminal instance of 4-coloring can be solved in
$O^*\!\left(
24^{s_{\star 1}/4}18^{n_{k3}}24^{n_{k4}}
\right)$ time.
\end{lemma}

\subsection{Terminal Algorithms via Path Decomposition DP}
We first enumerate the colors of all vertices in the two star
packings. The first stars have at most $24^{s_{\star 1}/4}$ colorings.
For a clique $C$ containing $i$ centers of the second star
packing, we assign distinct colors to these centers.
Let $Q_i$ denote an upper bound on the number of colorings
of the $i$ centers in $C$ and their leaves.
The centers must receive distinct colors. If none receives
color $4$, their leaves have at most $8^i$ colorings;
if one receives color $4$, their leaves have at most
$27\cdot 8^{i-1}$ colorings. Thus, for $i\in[4]$,
\[
Q_i
=\frac{6(4-i)}{(4-i)!}8^i
+\frac{6i}{(4-i)!}27\cdot8^{i-1}
=\frac{6\cdot8^i}{(4-i)!}
 \left(4-i+\frac{27i}{8}\right).
\]
We also set $Q_0=1$.
Thus, the vertices in
$S_{\star 1}\cup U_{\star 2}\cup S_{\star 2}$
have at most $\mathcal Q$ colorings, where
\begin{equation} \label{eq:four-color-stars-cost}
\mathcal Q
=24^{s_{\star 1}/4}
 \prod_{i=0}^{3}Q_i^{a_i}
 \prod_{i=0}^{4}Q_i^{b_i}.
\end{equation}

Let $X=\{v\in S'\mid 4\le|N_G(v)\cap(U\setminus U_{\star 2})|\}$.
We enumerate the colors of $X$ in at most $3^{|X|}$ ways.
This leaves a list 4-coloring instance on
$J=G[(S'\setminus X)\cup(U\setminus U_{\star 2})]$,
with lists obtained from $[3]$ for vertices in $S'\setminus X$
and $[4]$ for those in $U\setminus U_{\star 2}$,
by removing colors used by precolored neighbors.
By \cref{lm:four-color-star-maximality},
$\Delta(G[S'])\le2$, and every vertex of
$U\setminus U_{\star 2}$ has at most two neighbors in $S'$.
Hence, $\Delta(J)\le5$, and we apply
\cref{lm:pathwidth-DP} to solve this instance.

We bound the width of a path decomposition of $J$ using
\cref{lm:pathwidth-degree3-excess-bound}.
For each clique $C$ of $G[U]$, let
$d_C=|V(C)\setminus U_{\star 2}|$.
For $Y\subseteq S'$, let $E_{\le2}(Y)$ and $E_{\ge3}(Y)$
denote the numbers of edges from $Y$ to vertices in
$V(C)\setminus U_{\star 2}$, summed over cliques $C$
with $d_C\le2$ and $d_C\ge3$, respectively.
Since $\Delta(G[S'])\le2$ and
$\Delta(G[U\setminus U_{\star 2}])\le3$, with exactly
$4b_0$ vertices of degree three in the latter graph, we have
$
\sum_{v\in V(J)}[\deg_J(v)-2]_+
\le
4b_0+\frac32E_{\le2}(S'\setminus X)
     +2E_{\ge3}(S'\setminus X)
$.
The coefficient of $E_{\le2}(S'\setminus X)$ can be reduced
from $2$ to $3/2$ because each vertex in a remaining clique
of size at most two has internal degree at most one.
If it has $h\le2$ neighbors in $S'\setminus X$, its
contribution is at most $[h-1]_+\le h/2$.
Thus, \cref{lm:pathwidth-degree3-excess-bound} gives a
path decomposition of $J$ of width at most
$w+\varepsilon n+O(1)$, where
\[
w=\frac23b_0+\frac14E_{\le2}(S'\setminus X)
             +\frac13E_{\ge3}(S'\setminus X).
\]
Since $\Delta(J)\le5$, \cref{lm:pathwidth-DP} solves
list coloring of $J$ in randomized
$O^*(3^{w+\varepsilon n})$ time.

Since each $v\in X$ has at least four neighbors in
$U\setminus U_{\star 2}$, we have
$|X|\le\frac14(E_{\le2}(X)+E_{\ge3}(X))$.
Moreover, by \cref{lm:four-color-star-maximality}, each
$u\in U\setminus U_{\star 2}$ has at most two neighbors
in $S'$. Hence
\[
\begin{aligned}
|X|+w
&\le \frac23b_0+\frac14E_{\le2}(S')
                  +\frac13E_{\ge3}(S')\\
&\le \frac23b_0
    +\frac12\sum_{C:d_C\le2}d_C
    +\frac23\sum_{C:d_C\ge3}d_C\\
&= \frac{10}{3}b_0+2(a_0+b_1)
   +(a_1+b_2)+\frac12(a_2+b_3),
\end{aligned}
\]
where $\sum_{C:d_C\le2}$ denotes the sum over all cliques
$C$ of $G[U]$ with $d_C\le2$, and $\sum_{C:d_C\ge3}$
is defined analogously. The last equality uses
$\sum_{C:d_C\le2}d_C=2(a_1+b_2)+(a_2+b_3)$ and
$\sum_{C:d_C\ge3}d_C=3(a_0+b_1)+4b_0$.
Combining the at most $\mathcal Q$ colorings of
$S_{\star 1}\cup U_{\star 2}\cup S_{\star 2}$,
the at most $3^{|X|}$ colorings of $X$, and the
$O^*(3^{w+\varepsilon n})$-time list-coloring algorithm
for $J$, we obtain the following bound.

\begin{lemma} \label{lm:four-color-via-path-width}
Let $\mathcal Q$ be defined as in
\eqref{eq:four-color-stars-cost}.
For every fixed $\varepsilon>0$, there is a randomized
algorithm that solves a terminal instance of 4-coloring
in time
\[
O^*\!\left(
\mathcal Q
3^{\frac{10}{3}b_0+2(a_0+b_1)+(a_1+b_2)
+\frac12(a_2+b_3)+\varepsilon n}
\right).
\]
\end{lemma}

\subsection{From Terminal Bounds to the Overall Running Time}
\label{sec:four-color-terminal-to-overall}
We describe how the branching reductions of \cite{wu_et_al:LIPIcs.ESA.2024.103} can be used to derive an overall running-time bound from the running times for terminal instances.
Recall that for an instance $P=(G,I,S)$, $n_{k3}$ and $n_{k4}$ denote the numbers of $K_3$-components and $K_4$-components of $G[U]$, respectively.
Let $\mathcal{C}(P)$ denote the set of all other components of $G[U]$.
Following the Equation~(2) of \cite{wu_et_al:LIPIcs.ESA.2024.103}, we define the measure
\[
\mu(P)
=\omega_{c3}|S|+\omega_{k3}n_{k3}+\omega_{k4}n_{k4}
+\sum_{C\in\mathcal{C}(P)}
\sum_{v\in V(C)}\omega_{\deg_C(v)},
\]
where $\omega_{c3},\omega_{k3},\omega_{k4}$ and $\omega_i$ for $i\ge 0$ are nonnegative weights, with $\omega_0=0$ and $\omega_i=1$ for all $i\ge 3$.

The constraints on the weights in \cite[Equation~(3)]{wu_et_al:LIPIcs.ESA.2024.103} ensure that the measure does not increase under the reduction rules.
The branching rules for nonterminal instances are described in \cite[Section~3.2]{wu_et_al:LIPIcs.ESA.2024.103}, and the additional constraints in \cite[Equations~(6)--(21)]{wu_et_al:LIPIcs.ESA.2024.103} ensure that, for a target running-time base $r>1$, the total running-time budget of the resulting subinstances does not exceed that of the original instance.

\begin{lemma}[\cite{wu_et_al:LIPIcs.ESA.2024.103}]
\label{lm:four-color-branching}
Suppose that $r>1$ and the measure weights satisfy
the constraints in \cref{app:four-color-branching-constraints},
taken from Equations~(3) and~(6)--(21) of
\cite{wu_et_al:LIPIcs.ESA.2024.103}.
For every nonterminal instance $P=(G,I,S)$, the branching reductions in Section~3.2 of \cite{wu_et_al:LIPIcs.ESA.2024.103} produce subinstances $P_j=(G,I_j,S_j)$ for $j\in[t]$ such that
$\sum_{j=1}^t r^{\mu(P_j)}\le r^{\mu(P)}$.
Moreover, $P$ admits a valid 4-coloring if and only if at least one of the subinstances does, and $|U_j|<|U|$ holds for every $j\in[t]$, where $U_j=V(G)\setminus(I_j\cup S_j)$.
These subinstances can be generated with polynomial overhead per subinstance.
\end{lemma}

We next use a linear programming duality argument to combine
multiple algorithms for terminal instances with the branching
reductions.
Our formulation expresses $|S|$, $n_{k3}$, and $n_{k4}$ as linear functions of finer structural parameters.
This leads to a convex optimization formulation for the
overall running-time bound.

\begin{lemma}
\label{lm:four-color-terminal-combination}
Fix a constant $\gamma\ge0$.
Let $r>1$ and suppose that $r$ and the measure weights satisfy
the constraints in \cref{app:four-color-branching-constraints}.
Let $n_{\mathrm{par}}$ and $n_{\mathrm{alg}}$ be the numbers
of parameters and terminal algorithms, respectively.
Suppose every terminal instance has a polynomial-time
computable parameter vector
$\mathbf{x}\in\mathbb{R}_{\ge0}^{n_{\mathrm{par}}}$ satisfying
\begin{equation}
\mathbf M\mathbf x=
\begin{pmatrix}
|S| & n_{k3} & n_{k4}
\end{pmatrix}^{\top}
\end{equation}
for a fixed matrix
$\mathbf M\in\mathbb R^{3\times n_{\mathrm{par}}}$.
For each row $\mathbf R_i$ of a fixed matrix
$\mathbf R\in\mathbb R^{n_{\mathrm{alg}}\times n_{\mathrm{par}}}$,
suppose an algorithm solves every terminal instance in
$O^*(e^{\mathbf R_i\mathbf x+\gamma n})$ time,
where $n$ is the number of vertices of the original input.
If a nonnegative vector
$\boldsymbol\eta\in\mathbb R_{\ge0}^{n_{\mathrm{alg}}}$
satisfies
\begin{equation}\label{eq:four-color-dual}
(\ln r)
\begin{pmatrix}
\omega_{c3}&\omega_{k3}&\omega_{k4}
\end{pmatrix}
\mathbf M
-\boldsymbol\eta^{\top}\mathbf R
\ge\mathbf0^{\top},
\qquad
\mathbf1^{\top}\boldsymbol\eta=1,
\end{equation}
then 4-coloring can be solved in
$O^*((re^\gamma)^n)$ time.
\end{lemma}

\begin{proof}
For a terminal instance $P$ with parameter vector
$\mathbf x$, we have
\[
\begin{aligned}
\min_{i\in[n_{\mathrm{alg}}]}
  \bigl(\mathbf R_i\mathbf x+\gamma n\bigr)
&\le \boldsymbol\eta^{\top}\mathbf R\mathbf x+\gamma n\\
&\le (\ln r)\bigl(
\omega_{c3}|S|
+\omega_{k3}n_{k3}
+\omega_{k4}n_{k4}
\bigr)+\gamma n\\
&= (\ln r)\mu(P)+\gamma n.
\end{aligned}
\]
The first inequality follows from
$\boldsymbol\eta\ge\mathbf0$ and
$\mathbf1^{\top}\boldsymbol\eta=1$,
and the second follows from \eqref{eq:four-color-dual}
and $\mathbf x\ge\mathbf0$.
The equality holds because every component of
$G[U]$ in a terminal instance is $K_3$ or $K_4$.
Thus, choosing the terminal algorithm with the smallest
bound yields a running time of
$O^*(e^{\gamma n}r^{\mu(P)})$.

For nonterminal instances, we apply
\cref{lm:four-color-branching}.
Let $P_{\mathrm{root}}=(G,\emptyset,\emptyset)$ be the
initial instance, and let $\mathcal L_{\mathrm{term}}$
denote the multiset of terminal instances at the leaves
of the resulting branching tree.
By repeatedly applying the inequality in
\cref{lm:four-color-branching}, we obtain
\[
\sum_{L\in\mathcal L_{\mathrm{term}}}r^{\mu(L)}
\le r^{\mu(P_{\mathrm{root}})}.
\]
Consequently, the total time spent solving terminal
instances is
\[
O^*\!\left(
e^{\gamma n}
\sum_{L\in\mathcal L_{\mathrm{term}}}r^{\mu(L)}
\right)
=
O^*\!\left(e^{\gamma n}r^{\mu(P_{\mathrm{root}})}\right).
\]

Since $|U|$ strictly decreases in every reduction,
the depth of the branching tree is at most $n$.
Moreover, the inequality in \cref{lm:four-color-branching}
and the nonnegativity of $\mu$ imply that the number
of nodes at each depth is at most
$r^{\mu(P_{\mathrm{root}})}$.
Thus, the polynomial overhead of the reductions is
absorbed into the $O^*(\cdot)$ notation.

Finally, the constraints in \eqref{eq:wu-esa2024-eq-3},
together with $\omega_3=1$, imply
$\omega_1,\omega_2\le1$, $\omega_{k3}\le3$, and
$\omega_{k4}\le4$, yielding $\mu(P_{\mathrm{root}})\le n$.
Therefore, the overall running time is $O^*(e^{\gamma n}r^n)=O^*((re^\gamma)^n)$.
\end{proof}

Following the rescaling approach in \cite[Appendix~A]{wu_et_al:LIPIcs.ESA.2024.103}, we introduce $z=\ln r$ and scaled weights $\widehat\omega_j=z\omega_j$.
Under this substitution, the constraint in \eqref{eq:four-color-dual} becomes linear, while the branching constraints \eqref{eq:wu-esa2024-eq-6-21} become convex, as they take the form $\sum_j e^{f_j}\le1$ for affine functions $f_j$.

The terminal algorithm described in \cite[Section~3.1]{wu_et_al:LIPIcs.ESA.2024.103} runs in $O^*(\Lambda^{|S|}4.837^{n_{k3}}6^{n_{k4}})$ time.
Thus, their overall $O^*(1.7159^n)$ bound for 4-coloring can be recovered using \cref{lm:four-color-terminal-combination} by setting
\[
\mathbf{x}=
\begin{pmatrix}
|S| & n_{k3} & n_{k4}
\end{pmatrix}^\top,
\quad
\mathbf{M}=\mathbf{I}_3,
\quad
\mathbf{R}=
\begin{pmatrix}
\ln \Lambda & \ln 4.837 & \ln 6
\end{pmatrix},
\quad
\gamma=0,
\]
where $\mathbf{I}_3$ denotes the $3\times3$ identity matrix, $n_{\mathrm{par}}=3$, and $n_{\mathrm{alg}}=1$.
Hence, their convex optimization is a special case of our
formulation using $|S|$, $n_{k3}$, and $n_{k4}$ directly
as parameters, with a single terminal algorithm.

We now apply this formulation to our terminal algorithms.
We use the following $11$ parameters:
\[
\mathbf{x}=
\begin{pmatrix}
s_{\star 1}&s'&
a_0&a_1&a_2&a_3&
b_0&b_1&b_2&b_3&b_4
\end{pmatrix}^{\top}.
\]
From \eqref{eq:four-color-parameters},
the corresponding matrix is
\[
\mathbf M=
\begin{pmatrix}
1&1&0&3&6&9&0&3&6&9&12\\
0&0&1&1&1&1&0&0&0&0&0\\
0&0&0&0&0&0&1&1&1&1&1
\end{pmatrix}.
\]

Each terminal algorithm gives one row of $\mathbf R$
by expressing its leading exponent as a linear function
of $\mathbf x$.
We fix $\varepsilon=10^{-9}$ and
$\gamma=\varepsilon\ln3$.
Then the factor $3^{\varepsilon n}$ in
\cref{lm:four-color-via-path-width} equals $e^{\gamma n}$,
and the factor $e^{o(n)}$ in
\cref{lm:four-color-via-csp} is at most $e^{\gamma n}$
for sufficiently large $n$.
Thus, each terminal running-time bound in
\cref{lm:four-color-via-three-color,lm:four-color-via-csp,lm:four-color-via-tree-width,lm:four-color-via-path-width}
has the form $O^*(e^{\mathbf R_i\mathbf x+\gamma n})$
required by \cref{lm:four-color-terminal-combination}.
Here, we use the 3-coloring algorithms in
\cref{thm:ThreeDet,thm:ThreeRand} when applying
\cref{lm:four-color-via-three-color} in the deterministic
and randomized analyses, respectively.

The explicit matrices $\mathbf R$ for the deterministic
and randomized algorithms, together with dual certificates
for the resulting bounds, are given in
\cref{app:four-coloring-optimization}.
Here we state the results.
Let $r_{\mathrm{det}}$ and $r_{\mathrm{rand}}$
denote the optimal values of $r$ in the
deterministic and randomized cases, respectively.
We obtain
$r_{\mathrm{det}} e^\gamma\le\FourDetBase$ and
$r_{\mathrm{rand}} e^\gamma\le\FourRandBase$,
which yield the following theorems.

\ThmFourDet*
\ThmFourRand*

\section*{AI methodology}
We used ChatGPT to assist in generating code for finding
feasible solutions, computing upper bounds on the optimal
values of the linear and convex programs, and obtaining
corresponding dual certificates.
This assistance materially affected
\cref{app:three-coloring-LP,app:four-coloring-optimization,app:meijer-lp}.
The authors verified the correctness and originality of all content.


\bibliographystyle{alphaurl}
\bibliography{references}

\appendix

\section{Decision-to-Search Reduction}
\label{app:decision-to-search}

A standard self-reduction allows us to construct a
$k$-coloring using a decision algorithm with only
polynomial overhead.
This reduction also appears in the survey
\cite{Husfeldt_2015}.

\begin{lemma}
\label{lm:decision-to-search}
Suppose that the existence of a $k$-coloring of an
$n$-vertex graph $G$ can be decided in $O(\tau(n))$ time.
Then a $k$-coloring of $G$ can be constructed, if one exists,
in $O^*(\tau(n))$ time.
\end{lemma}

\begin{proof}
We first use the decision algorithm to check whether
$G$ admits a $k$-coloring.
If not, we report that no such coloring exists.
Otherwise, we repeatedly consider nonadjacent vertices
$u,v$ and test whether adding $uv$ preserves
$k$-colorability.
If so, we add $uv$ to $G$.
We stop when no further edge can be added.

A graph is \emph{complete multipartite} if its vertex set
can be partitioned into independent sets such that
every two vertices in different parts are adjacent.
The resulting graph is complete multipartite with at most
$k$ parts, since otherwise some nonadjacent vertices
could receive different colors in a $k$-coloring,
allowing another edge to be added.
Thus, its complement is a disjoint union of at most
$k$ cliques, which yield a $k$-coloring of the original graph.
Since $O(n^2)$ decision calls are needed,
the total running time is $O^*(\tau(n))$.
\end{proof}

\section{Deferred Proofs}
\subsection{Proof of Lemma~\ref{lm:pathwidth-degree3-excess-bound}}
\label{app:treewidth-degree3-excess-bound}

Here, we restate \cref{lm:pathwidth-upperbound} for convenience.
\LmPathWidthUpperbound*

\LmPathWidthDegreeExcessBound*
\begin{proof}
Let $G'$ be obtained from $G$ by replacing each degree-$7$
vertex $v$ with two adjacent vertices $v_1$ and $v_2$ of
degrees $4$ and $5$, respectively.
This is done by assigning three neighbors of $v$ to $v_1$
and the remaining four to $v_2$.

Apply \cref{lm:pathwidth-upperbound} to $G'$ with
$\varepsilon/2$.
For every degree $d\neq 7$, the coefficient in
\cref{lm:pathwidth-upperbound} is at most $[d-2]_+/6$.
This follows directly for $d\le 6$, and for $d\ge 8$
from $1\le(d-2)/6$.
For each split degree-$7$ vertex, the two new vertices
contribute
\[
\frac13+\frac{13}{30}
=\frac{23}{30}
<\frac56
=\frac{7-2}{6}.
\]
Hence, the degree-weighted part of the pathwidth bound
for $G'$ is at most
$(\sum_{v\in V(G)}[\deg_G(v)-2]_+)/6$.
Since $n'=|V(G')|\le 2n$, the remaining terms satisfy
$\frac{\varepsilon}{2}n'+O(1)
\le \varepsilon n+O(1)$.
Together with the bound on the degree-weighted part, this gives the desired pathwidth bound for $G'$.

Finally, contracting each edge $v_1v_2$ recovers $G$.
The nodes whose bags contain $v_1$ and $v_2$,
respectively, induce subpaths of $\mathcal{T}$.
Since $v_1v_2$ is an edge, these subpaths intersect,
and hence their union is also a subpath.
Thus, replacing $v_1$ and $v_2$ by $v$ in every bag
preserves the path-decomposition properties without
increasing any bag size.
The resulting path decomposition of $G$ satisfies
the claimed bound and can be constructed in polynomial time.
\end{proof}

\subsection{Proof of Lemma~\ref{lm:bushy-forest-normalize}}\label{app:lemma-bushy-forest-normalize}

\LmBushyForestNormalize*

\begin{proof}
Suppose that there exists a vertex $v\notin V(F)$ with
$\deg_{G-V(F)}(v)=3$ that is adjacent to a leaf $\ell$ of a tree
$B$ in $F$, where $B$ is not a $K_{1,4}$.
Let $u_1,u_2,u_3$ denote the three neighbors of $v$ in $G-V(F)$.
We distinguish two cases according to the number of internal vertices
of $B$ and show how to modify $F$ to eliminate this violating
configuration.

\textbf{Case 1: $B$ has at least two internal vertices.}
Since $B$ has at least two internal vertices, there exists an internal
vertex $i$ of $B$ that is not adjacent to $\ell$ in $B$.
We remove $B$ from $F$ and add a new tree consisting of $i$ and all
of its neighbors in $B$, with $i$ as its only internal vertex.
Since $\deg_B(i)\ge 4$, this tree satisfies the bushy-forest condition.
We also add a new $K_{1,4}$ centered at $v$ with leaves
$\ell,u_1,u_2,u_3$.

\textbf{Case 2: $B$ has exactly one internal vertex.}
Since $B$ is not a $K_{1,4}$, its internal vertex has at least five
leaves. We remove $\ell$ from $B$; the remaining tree still satisfies
the bushy-forest condition, since its internal vertex has at least four
leaves. We then add to $F$ a new $K_{1,4}$ centered at $v$ with leaves
$\ell,u_1,u_2,u_3$.

In either case, the resulting forest is still a bushy forest, the chosen
violating configuration is eliminated, and the number of trees in the
forest increases by one. The resulting forest may no longer be maximal,
so we greedily extend it.
This extension preserves the bushy-forest condition and never decreases
the number of trees.

We repeat this modification and extension as long as a violating
configuration exists. Each modification strictly increases the number
of trees in $F$, while the greedy extension never decreases it.
Since the number of trees in the forest is bounded above by
$|V(G)|=n$, at most $n$ modifications are performed.
Each modification and greedy extension can be carried out in polynomial
time. Hence, the entire procedure runs in polynomial time and terminates
with a maximal bushy forest satisfying the desired property.
\end{proof}

\section{Linear Programs and Dual Certificates for 3-Coloring}
\label{app:three-coloring-LP}

We formulate the linear programs used to derive the
running-time bounds in \cref{thm:ThreeDet,thm:ThreeRand}.
Throughout this section, we use $p,q,r,s,t,d,c,m$
for the corresponding parameters divided by $n$,
and fix $\varepsilon=10^{-9}$.
We set $\kappa=3$ for the deterministic analysis
and $\kappa=2$ for the randomized analysis.
The normalized logarithms of the running-time bounds
in \cref{lm:three-coloring-csp-packing,lm:three-coloring-subcubic-width-DP,lm:three-coloring-treewidth-DP}
are
\[
\begin{aligned}
L_{\mathrm{csp}}
={}&p\ln3+q\ln2+(s+t)\ln\Lambda
+\left(\frac{d+t}{10}+\frac{19c}{660}\right)\ln(3\Lambda^{-4}),\\
L_{\mathrm{tw}}
={}&p\ln3+q\ln2+m\ln3
+\left(\frac{d+t+2r-4m}{6}+\varepsilon\right)\ln3,\\
L_{\mathrm{pw}}(\kappa)
={}&p\ln3+q\ln2+m\ln3+(r-2m)\ln2
+\left(\frac{c}{6}+\varepsilon\right)\ln\kappa.
\end{aligned}
\]

The structural constraints are
\[
\begin{aligned}
&p,q,r,s,t,d,c,m\ge0,\qquad
p+q+r+s+t=1,\qquad d\le s,\\
&4p+2q\le r,\qquad s\le2r,\qquad
d\le8p,\qquad t\le2s+d,\qquad
c\le d+t,\qquad 2m\le r.
\end{aligned}
\]

For each $\kappa\in\{2,3\}$, the primal linear program
maximizes $\tau$ subject to these structural constraints and
\[
\tau\le L_{\mathrm{csp}},\qquad
\tau\le L_{\mathrm{tw}},\qquad
\tau\le L_{\mathrm{pw}}(\kappa).
\]
In the randomized analysis, we omit the deterministic bound
$L_{\mathrm{pw}}(3)$ since it is dominated by
$L_{\mathrm{pw}}(2)$.

\paragraph{Dual linear programs.}
Let $\mathbf{x}=(p,q,r,s,t,d,c,m)^\top
\in\mathbb{R}_{\ge0}^8$.
We express the structural constraints as
$\mathbf{A}\mathbf{x}\le\mathbf{0}$
and $\mathbf{u}^\top\mathbf{x}=1$, where
$\mathbf{A}\in\mathbb{R}^{7\times8}$ and
$\mathbf{u}\in\mathbb{R}^8$ are given by
\[
\mathbf{A}=
\begin{pmatrix}
 0&0& 0&-1& 0& 1& 0& 0\\
 4&2&-1& 0& 0& 0& 0& 0\\
 0&0&-2& 1& 0& 0& 0& 0\\
-8&0& 0& 0& 0& 1& 0& 0\\
 0&0& 0&-2& 1&-1& 0& 0\\
 0&0& 0& 0&-1&-1& 1& 0\\
 0&0&-1& 0& 0& 0& 0& 2
\end{pmatrix},
\quad
\mathbf{u}=
\begin{pmatrix}
1\\1\\1\\1\\1\\0\\0\\0
\end{pmatrix}.
\]

We express the three normalized logarithmic
running-time bounds as
$\mathbf{B}_\kappa\mathbf{x}+\mathbf{b}_\kappa$,
where $\mathbf{B}_\kappa\in\mathbb{R}^{3\times8}$
and $\mathbf{b}_\kappa\in\mathbb{R}^3$ are given by
\[
\mathbf{B}_\kappa=
\begin{pmatrix}
\ln3 & \ln2 & 0 & \ln\Lambda
& \ln\Lambda+\frac{\ln(3\Lambda^{-4})}{10}
& \frac{\ln(3\Lambda^{-4})}{10}
& \frac{19\ln(3\Lambda^{-4})}{660} & 0\\
\ln3 & \ln2 & \frac{\ln3}{3} & 0
& \frac{\ln3}{6} & \frac{\ln3}{6}
& 0 & \frac{\ln3}{3}\\
\ln3 & \ln2 & \ln2 & 0
& 0 & 0 & \frac{\ln\kappa}{6}
& \ln3-2\ln2
\end{pmatrix},
\]
and
\[
\mathbf{b}_\kappa=
\begin{pmatrix}
0 & \varepsilon\ln3 & \varepsilon\ln\kappa
\end{pmatrix}^\top.
\]
The rows of $\mathbf{B}_\kappa$ correspond to
the algorithms via CSP, treewidth-DP, and pathwidth-DP,
respectively, while its columns correspond to
$p,q,r,s,t,d,c,m$.

Let $\mathbf{1}\in\mathbb{R}^3$
denote the all-ones vector.
For each $\kappa\in\{2,3\}$,
the primal linear program is
\[
\begin{array}{ll}
\text{maximize}
&\tau\\
\text{subject to}
&\tau\mathbf{1}
 \le\mathbf{B}_\kappa\mathbf{x}+\mathbf{b}_\kappa,\\
&\mathbf{A}\mathbf{x}\le\mathbf{0},\\
&\mathbf{u}^\top\mathbf{x}=1,\\
&\mathbf{x}\ge\mathbf{0},
 \quad\tau\in\mathbb{R}.
\end{array}
\]

Its dual is
\[
\begin{array}{ll}
\text{minimize}
&z+\boldsymbol{\eta}^\top \mathbf{b}_\kappa \\
\text{subject to}
&\boldsymbol{\eta}^\top\mathbf{B}_\kappa \le
z\mathbf{u}^\top
+\boldsymbol{\xi}^\top\mathbf{A},\\
&\mathbf{1}^\top\boldsymbol{\eta}=1,\\
&\boldsymbol{\eta}\in
 \mathbb{R}_{\ge0}^3,\quad
 \boldsymbol{\xi}\in
 \mathbb{R}_{\ge0}^7,\quad
 z\in\mathbb{R}.
\end{array}
\]

Here, $\boldsymbol{\eta}$ and
$\boldsymbol{\xi}$ correspond to the
running-time constraints and structural
inequalities, respectively, while $z$
corresponds to the equality
$p+q+r+s+t=1$.

By weak duality, any feasible primal and dual solutions
satisfy
\[
\begin{aligned}
\tau
&\le \boldsymbol{\eta}^\top
(\mathbf{B}_\kappa\mathbf{x}+\mathbf{b}_\kappa)\\
&\le (z\mathbf{u}^\top+\boldsymbol{\xi}^\top\mathbf{A})
\mathbf{x}+\boldsymbol{\eta}^\top\mathbf{b}_\kappa\\
&\le z+\boldsymbol{\eta}^\top\mathbf{b}_\kappa.
\end{aligned}
\]

\paragraph{Feasible dual certificates.}
We give feasible dual solutions for $\kappa=3$ and
$\kappa=2$ in the following table.
All displayed decimal values are treated as exact
rational numbers.
\[
\begin{array}{c|cc}
&\kappa=3&\kappa=2\\ \hline
z
&0.277893717&0.274458566\\
\eta_{\mathrm{csp}}
&0.872081561&0.881461989\\
\eta_{\mathrm{tw}}
&0.108128455&0.052151678\\
\eta_{\mathrm{pw}}
&0.019789984&0.066386333\\
\xi_1
&0.006293660&0.000000010\\
\xi_2
&0.207626932&0.209344508\\
\xi_3
&0&0\\
\xi_4
&0.001223624&0.001110933\\
\xi_5
&0.000305899&0.000277726\\
\xi_6
&0.000000010&0.004006697\\
\xi_7
&0.016951940&0.000000010
\end{array}
\]

For both values of $\kappa$, these solutions satisfy
$\mathbf{1}^\top\boldsymbol{\eta}=1$,
$\boldsymbol{\eta}\ge\mathbf{0}$, and
$\boldsymbol{\xi}\ge\mathbf{0}$.
They also satisfy the componentwise inequalities
\[
\boldsymbol{\eta}^\top\mathbf{B}_\kappa
\le z\mathbf{u}^\top+\boldsymbol{\xi}^\top\mathbf{A}.
\]
The corresponding dual objective values satisfy
\[
\begin{aligned}
\kappa=3:\quad
z+\boldsymbol{\eta}^\top\mathbf{b}_3
&\le0.277893718\le\ln1.3204,\\
\kappa=2:\quad
z+\boldsymbol{\eta}^\top\mathbf{b}_2
&\le0.274458567\le\ln1.3159.
\end{aligned}
\]
For numerical verification, we used rational interval bounds
for $\ln$ to verify both the componentwise inequalities
and the objective bounds.
These checks establish the feasibility of the displayed
dual solutions and the stated objective bounds independently
of the accuracy of the numerical optimizer.
By weak duality, the certificates establish the claimed
deterministic and randomized running-time bounds.

\section{Branching Constraints for 4-Coloring}\label{app:four-color-branching-constraints}

We restate \cite[Equation~(3)]{wu_et_al:LIPIcs.ESA.2024.103}
as follows:
\begin{equation} \label{eq:wu-esa2024-eq-3}
\left\{
\begin{aligned}
&0\le\omega_{c3}\le\omega_1\le\omega_2\le\omega_3,\\
&\omega_3-\omega_2\le\omega_2-\omega_1\le\omega_1,\\
&3\omega_2\ge\omega_{k3},\\
&4\omega_3\ge\omega_{k4}.
\end{aligned}
\right.
\end{equation}

For \cite[Equations~(6)--(21)]{wu_et_al:LIPIcs.ESA.2024.103},
each row of
\cref{tab:wu-branching-constraints} specifies a multiset
$\mathcal{D}$ of pairs $(\alpha,\beta)$, representing the constraint
\begin{equation} \label{eq:wu-esa2024-eq-6-21}
\sum_{(\alpha,\beta)\in\mathcal{D}} \alpha r^{-\beta}\le1.
\end{equation}
We write $\delta_\omega=\omega_2-\omega_{c3}$ and
$M_j=|\mathcal{MI}(C_j)|$, where $\mathcal{MI}(C_j)$
denotes the family of all maximal independent sets of
the cycle $C_j$, and
$(M_4,M_5,M_6,M_7,M_8)=(2,5,5,7,10)$.

\begin{table}[t]
\centering
\small
\renewcommand{\arraystretch}{1.15}
\begin{tabular}{@{}cl@{}}
\toprule
\cite{wu_et_al:LIPIcs.ESA.2024.103} Eq. & Pairs $(\alpha,\beta)$ in $\mathcal{D}$\\
\midrule
(6)
& $(2,2\omega_1-\omega_{c3})$\\
(7a)
& $(M_j,j\delta_\omega+\lceil j/3\rceil\omega_{c3})$
  \quad for $j\in\{4,5,6,7,8\}$\\
(7b)
& $(1,3\delta_\omega+\omega_{c3})$,
  $(1,4\delta_\omega+\omega_{c3})$,
  $(1,5\delta_\omega+\omega_{c3})$\\
(8a)
& $(1,1+2\delta_\omega)$,
  $(2,1+3\delta_\omega)$\\
(8b)
& $(2,1+2\delta_\omega)$,
  $(1,1+4\delta_\omega)$\\
(8c)
& $(1,2+\delta_\omega-\omega_{c3})$,
  $(1,1+3\delta_\omega)$,
  $(1,2+3\delta_\omega-\omega_{c3})$\\
(8d)
& $(1,2+\delta_\omega-\omega_{c3})$,
  $(1,1+3\delta_\omega)$,
  $(1,2+2\delta_\omega-\omega_{c3})$\\
(9)
& $(1,2+5\delta_\omega+\omega_{c3}-2\omega_1)$,
  $(1,2+2\delta_\omega+\omega_{c3}-\omega_{k3})$\\
(10)
& $(1,2+4\delta_\omega+\omega_{c3}-2\omega_1)$,
  $(2,2+4\delta_\omega+2\omega_{c3}-\omega_{k3})$\\
(11)
& $(1,4+2\delta_\omega-\omega_{c3}-2\omega_1)$,
  $(1,2+2\delta_\omega-\omega_{c3})$,
  $(2,3+4\delta_\omega-2\omega_1)$\\
(12)
& $(1,1+5\delta_\omega)$,
  $(1,1-\omega_{c3})$\\
(13), (14)
& $(1,4-3\omega_{c3})$,
  $(3,5-4\omega_{c3})$\\
(15)
& $(1,4-3\omega_{c3})$,
  $(1,5-\omega_{c3}-\omega_{k4})$\\
(16)
& $(2,5-4\omega_{c3})$,
  $(1,6-2\omega_{c3}-\omega_{k4})$\\
(17a), (18b)
& $(2,4-3\omega_{c3})$,
  $(1,5-4\omega_{c3})$,
  $(1,6-4\omega_{c3})$\\
(17b)
& $(2,4-3\omega_{c3})$,
  $(1,5-4\omega_{c3})$,
  $(2,8-6\omega_{c3})$\\
(18a)
& $(3,4-3\omega_{c3})$\\
(18c)
& $(3,4-3\omega_{c3})$,
  $(1,7-5\omega_{c3})$\\
(19)
& $(2,4-3\omega_{c3}+6(1-\omega_2))$,
  $(1,5-4\omega_{c3}+3(1-\omega_2))$,
  $(1,6-5\omega_{c3})$\\
(20)
& $(3,5-4\omega_{c3})$,
  $(1,6-5\omega_{c3})$,
  $(1,7-5\omega_{c3})$\\
(21)
& $(2,5-4\omega_{c3})$,
  $(2,6-5\omega_{c3})$,
  $(1,7-6\omega_{c3})$\\
\bottomrule
\end{tabular}
\caption{Branching constraints from
\cite[Equations~(6)--(21)]{wu_et_al:LIPIcs.ESA.2024.103}.}
\label{tab:wu-branching-constraints}
\end{table}

\section{Convex Programs and Dual Certificates for 4-Coloring}
\label{app:four-coloring-optimization}

We give the terminal coefficient matrices and feasible
solutions certifying our running-time bounds.
The parameter order is
\[
\mathbf x=
\begin{pmatrix}
s_{\star 1}&s'&
a_0&a_1&a_2&a_3&
b_0&b_1&b_2&b_3&b_4
\end{pmatrix}^{\top}.
\]
The corresponding matrix mapping these parameters to
$|S|$, $n_{k3}$, and $n_{k4}$ is
\[
\mathbf M=
\begin{pmatrix}
1&1&0&3&6&9&0&3&6&9&12\\
0&0&1&1&1&1&0&0&0&0&0\\
0&0&0&0&0&0&1&1&1&1&1
\end{pmatrix}.
\]

\paragraph{Terminal coefficient matrices.}
For $m\in\{3,4\}$ and $0\le i\le m$, define
\[
B_{m,i}(\beta)
=\ln m+(m-1+3i)\ln\beta,
\qquad
P_{m,i}=\ln Q_i+\psi(m-i)\ln3,
\]
where $Q_i$ is defined in the analysis of
\cref{lm:four-color-via-path-width}, and
\[
\bigl(\psi(0),\psi(1),\psi(2),\psi(3),\psi(4)\bigr)
=\left(0,\frac12,1,2,\frac{10}{3}\right).
\]
The parameter $\beta$ denotes the running-time base of
the chosen 3-coloring algorithm.

The rows $\mathbf R_{\mathrm{3col}}(\beta)$,
$\mathbf R_{\mathrm{csp}}$, $\mathbf R_{\mathrm{tree}}$,
and $\mathbf R_{\mathrm{path}}$ are obtained from the
running-time bounds in
\cref{lm:four-color-via-three-color},
\cref{lm:four-color-via-csp},
\cref{lm:four-color-via-tree-width}, and
\cref{lm:four-color-via-path-width}, respectively.
They are
\[
\begin{aligned}
\mathbf R_{\mathrm{3col}}(\beta)
={}&\bigl(
\ln\beta,\ln\beta,
B_{3,0}(\beta),\ldots,B_{3,3}(\beta),
B_{4,0}(\beta),\ldots,B_{4,4}(\beta)
\bigr),\\
\mathbf R_{\mathrm{csp}}
={}&\bigl(
\tfrac14\ln3,\ln\Lambda,
\ln g_{3,0},\ldots,\ln g_{3,3},
\ln g_{4,0},\ldots,\ln g_{4,4}
\bigr),\\
\mathbf R_{\mathrm{tree}}
={}&\bigl(
\tfrac14\ln24,0,
\ln18,\ln18,\ln18,\ln18,
\ln24,\ln24,\ln24,\ln24,\ln24
\bigr),\\
\mathbf R_{\mathrm{path}}
={}&\bigl(
\tfrac14\ln24,0,
P_{3,0},\ldots,P_{3,3},
P_{4,0},\ldots,P_{4,4}
\bigr).
\end{aligned}
\]
Here, $g_{m,i}$ is defined in
\cref{lm:four-color-via-csp}.
Thus, the deterministic and randomized coefficient
matrices are
\[
\mathbf R_{\mathrm{det}}=
\begin{pmatrix}
\mathbf R_{\mathrm{3col}}(\ThreeDetBase)\\
\mathbf R_{\mathrm{csp}}\\
\mathbf R_{\mathrm{tree}}
\end{pmatrix},
\qquad
\mathbf R_{\mathrm{rand}}=
\begin{pmatrix}
\mathbf R_{\mathrm{3col}}(\ThreeRandBase)\\
\mathbf R_{\mathrm{csp}}\\
\mathbf R_{\mathrm{tree}}\\
\mathbf R_{\mathrm{path}}
\end{pmatrix}.
\]

We use the deterministic and randomized 3-coloring
running-time bounds from \cref{thm:ThreeDet,thm:ThreeRand},
respectively.
We fix $\varepsilon=10^{-9}$ and
$\gamma=\varepsilon\ln3$.
The additional factor $3^{\varepsilon n}$ in
\cref{lm:four-color-via-path-width} equals $e^{\gamma n}$,
and the factor $e^{o(n)}$ in
\cref{lm:four-color-via-csp} is at most $e^{\gamma n}$
for sufficiently large $n$.
Consequently, every terminal algorithm satisfies the
bound $O^*(e^{\mathbf R_i\mathbf x+\gamma n})$
required by \cref{lm:four-color-terminal-combination}.

\paragraph{Convex formulation.}
We introduce the rescaled variables
\[
z=\ln r,\qquad \widehat\omega_j=z\omega_j.
\]
In particular, $\widehat\omega_3=z$, since $\omega_3=1$.
The terminal coefficient inequalities become
\begin{equation} \label{eq:four-color-terminal-coef-constraints}
\begin{pmatrix}
\widehat\omega_{c3}&\widehat\omega_{k3}&\widehat\omega_{k4}
\end{pmatrix}\mathbf M
-\boldsymbol\eta^{\top}\mathbf R
\ge\mathbf0^{\top},
\qquad
\boldsymbol\eta\ge\mathbf0,\qquad
\mathbf1^{\top}\boldsymbol\eta=1.    
\end{equation}
We combine these with the branching constraints in
\cref{app:four-color-branching-constraints}, expressed
in the rescaled variables.
The constraints in \eqref{eq:wu-esa2024-eq-3} remain linear,
while each branching inequality of \eqref{eq:wu-esa2024-eq-6-21} becomes a sum of exponentials of affine functions bounded by one.
Thus, minimizing $z$ subject to these constraints
is a convex optimization problem.

We use $\mathbf R=\mathbf R_{\mathrm{det}}$ and
$\mathbf R=\mathbf R_{\mathrm{rand}}$ for the deterministic
and randomized cases, respectively.
The vector $\boldsymbol\eta$ certifies that the minimum
terminal exponent is bounded by the terminal measure,
as shown in \cref{lm:four-color-terminal-combination}.

\paragraph{Feasible solutions and verification.}
The following table specifies feasible solutions.
All displayed finite decimals are interpreted as exact
rational numbers.
The original variables are recovered by
$r=e^z$ and $\omega_j=\widehat\omega_j/z$, with $\omega_i=1$ for $i \ge 3$.
\[
\begin{array}{c|cc}
 & \text{Deterministic} & \text{Randomized}\\ \hline
z
&0.5291126816180365&0.5266716867458051\\
\widehat\omega_1
&0.4944314348773351&0.4928717407360532\\
\widehat\omega_2
&0.5291126816180365&0.5266716867458051\\
\widehat\omega_{c3}
&0.2957156198800068&0.2925962315974430\\
\widehat\omega_{k3}
&1.5714124563519097&1.5621902025983386\\
\widehat\omega_{k4}
&1.8810074111165460&1.9480191762881393\\ \hline
\eta_{\mathrm{3col}}
&0.0788417191536246&0.2043107963374502\\
\eta_{\mathrm{csp}}
&0.8811402533346064&0.7611219494728269\\
\eta_{\mathrm{tree}}
&0.0400180275117690&0.0069512075211995\\
\eta_{\mathrm{path}}
&\text{--}&0.0276160466685234
\end{array}
\]
The entries of $\boldsymbol\eta$ sum to exactly one
in both cases.

For numerical verification, we used rational interval bounds
for $\exp$ and $\ln$ to verify all constraints in
\eqref{eq:wu-esa2024-eq-3} and
\eqref{eq:wu-esa2024-eq-6-21},
and all terminal coefficient inequalities in
\eqref{eq:four-color-terminal-coef-constraints}.
These checks establish the feasibility of the displayed
solutions independently of the accuracy of the numerical
optimizer.

Let $z_{\mathrm{det}}$ and $z_{\mathrm{rand}}$ denote the values
of $z$ in the displayed deterministic and randomized feasible
solutions, respectively.
The resulting running-time bases satisfy
\[
e^{z_{\mathrm{det}}+\gamma}
\le1.697425486\le\FourDetBase,
\qquad
e^{z_{\mathrm{rand}}+\gamma}
\le1.693287132\le\FourRandBase.
\]
Thus, these certificates establish the claimed deterministic
and randomized running-time bounds.

\section{On the Linear Program in Meijer's Preprint}
\label{app:meijer-lp}

Meijer \cite{meijer20233coloringtimeo13217n} reports an
$O^*(1.3217^n)$-time algorithm for 3-coloring.
However, the linear program presented in the preprint
admits a feasible solution with an objective value
exceeding $\ln1.3217$.
We restate the linear program below and give
an explicit feasible solution.

We retain the notation of the preprint, but throughout
this section all cardinalities are normalized by $n$;
thus, for example, $|R|$ denotes the original
cardinality divided by $n$.
Using natural logarithms, the linear program in
Section~5.2 of the preprint is
\[
\begin{array}{ll}
\text{maximize}
&
|R|\ln3+|I|\ln2+|N^*|\ln1.36443+|U^*|\ln1.34004
\\[2pt]
\text{subject to}
&
4|R|+2|I|\le|L|,
\\
&
|N_1|+2|N_2|+|N_3|\le2|L|,
\\
&
|U'|\ge
\frac15|N_{3,5}|+\frac13|N_{3,6}|
+\frac57|N_{3,7}|+|N_{3,8}|,
\\
&
\displaystyle
\sum_{j=0}^{7}\frac{10-j}{8-j}|U_j|
\le
2|N_2|+3\sum_{j=1}^{8}|N_{3,j}|-3|U'|,
\\
&
|N^*|=|N|-\frac35|N_{3,5}|
-|N_{3,6}|-|N_{3,7}|-|N_{3,8}|,
\\
&
|U^*|=|U|+\frac35|N_{3,5}|+|N_{3,6}|,
\\
&
|R|+|I|+|L|+|N|+|U|=1,
\\
&
\displaystyle \sum_{k=1}^{3}|N_k|=|N|,
\\
&
\displaystyle \sum_{j=1}^{8}|N_{3,j}|=|N_3|,
\\
&
\displaystyle \sum_{j=0}^{7}|U_j|+|U'|=|U|,
\\
&
\displaystyle \sum_{j=1}^{8}\frac8j|N_{3,j}|\le8|R|,
\\
&
\text{all variables are nonnegative}.
\end{array}
\]
These are the constraints in
\cite[Equations~(1)--(4) and~(6)--(12)]{meijer20233coloringtimeo13217n},
after normalization.

Consider the assignment
\[
\begin{aligned}
|R|&=\frac5{113},
&
|L|&=\frac{20}{113},
\\
|N|=|N^*|&=\frac{40}{113},
&
|U|=|U^*|&=\frac{48}{113},
\\
|N_1|=|N_3|=|N_{3,4}|&=\frac{20}{113},
&
|U_0|&=\frac{48}{113},
\end{aligned}
\]
with all other variables equal to zero.
All equality constraints hold.
Moreover, each inequality constraint holds with equality:
\[
\begin{aligned}
4|R|+2|I|
&=\frac{20}{113}=|L|,\\
|N_1|+2|N_2|+|N_3|
&=\frac{40}{113}=2|L|,\\
|U'|
&=0
=\frac15|N_{3,5}|+\frac13|N_{3,6}|
+\frac57|N_{3,7}|+|N_{3,8}|,\\
\sum_{j=0}^{7}\frac{10-j}{8-j}|U_j|
&=\frac{60}{113}
=2|N_2|+3\sum_{j=1}^{8}|N_{3,j}|-3|U'|,\\
\sum_{j=1}^{8}\frac8j|N_{3,j}|
&=\frac{40}{113}=8|R|.
\end{aligned}
\]
Thus, this assignment is feasible.
Its objective value is
\[
\frac{
5\ln3+40\ln1.36443+48\ln1.34004
}{113}
\approx0.2829389919,
\]
whose exponential is
\[
\left(3^5\,1.36443^{40}\,1.34004^{48}\right)^{1/113}
>1.32702
>1.3217.
\]

Consequently, the optimum of the displayed maximization
problem exceeds $\ln1.3217$, so this linear program
does not establish the reported $O^*(1.3217^n)$ running-time bound.
However, this feasible solution does not necessarily imply
a lower bound on the algorithm's running time, since it
need not correspond to a realizable graph.
Additional structural constraints might exclude this solution,
and a different analysis might establish the reported bound.


\end{document}